\documentclass[11pt,a4paper]{article}

\usepackage{microtype}
\usepackage{amsmath, amsfonts,amssymb,amsthm,mathtools,nicefrac,nccmath,cases,physics}
\usepackage{mathrsfs}  
\usepackage{bm}
\usepackage{soul}
\usepackage{dsfont}
\usepackage{booktabs,multirow}
\usepackage{tabularx}
\usepackage[usenames,dvipsnames,table]{xcolor}
\usepackage{colortbl}
\usepackage{caption}
\usepackage{subcaption}
\usepackage{float}
\usepackage{hhline}
\usepackage{enumitem}
\usepackage{cite}
\numberwithin{equation}{section}
\numberwithin{table}{section}
\usepackage[linktocpage=true]{hyperref}
\hypersetup{colorlinks=true,linkcolor=colorloc3,citecolor=colorloc3,urlcolor=colorloc3}
\usepackage{afterpage}

\def\hybrid{\topmargin -20pt    \oddsidemargin 0pt
	\headheight 0pt \headsep 0pt
	\textwidth 6.5in        % US paper
	\textheight 9in         % US paper
	\textwidth 6.25in       % A4 paper
	\textheight 9 in       % A4 paper
	\marginparwidth .875in
	\parskip 5pt plus 1pt 
	\jot = 1.5ex
}
\hybrid

\definecolor{colorloc1}{RGB}{164,42,46} % Pompeian red
\definecolor{colorloc2}{RGB}{100,100,100} % grey
\definecolor{colorloc3}{RGB}{204,119,34}  % Roman ochre
\definecolor{colorloc4}{RGB}{25,25,112}  % midnight blue
\definecolor{colorloc5}{RGB}{100,0,0}  % dark red
\definecolor{colorloc6}{RGB}{200,200,200}  % light gray
\definecolor{colorloc7}{RGB}{70,70,70}  % darker gray
\definecolor{colorloc8}{RGB}{0,128,128}  % teal

\usepackage[framemethod=default]{mdframed}

\newmdenv[skipabove=10pt,
skipbelow=7pt,
rightline=false,
leftline=true,
topline=false,
bottomline=false,
linecolor=colorloc4,
backgroundcolor=colorloc8!5,
innerleftmargin=4pt,
innerrightmargin=0pt,
innertopmargin=0pt,
leftmargin=2pt,
rightmargin=0pt,
linewidth=2pt,
innerbottommargin=4pt,
frametitlebackgroundcolor=colorloc3]{lbBox}
\newenvironment{importantbox}{\begin{lbBox}\vspace{3 mm}
	} {\vspace{1.5 mm}\end{lbBox}}

\newmdenv[skipabove=10pt,
skipbelow=7pt,
rightline=false,
leftline=true,
topline=false,
bottomline=false,
linecolor=colorloc5,
backgroundcolor=colorloc1!5,
innerleftmargin=4pt,
innerrightmargin=0pt,
innertopmargin=0pt,
leftmargin=2pt,
rightmargin=0pt,
linewidth=2pt,
innerbottommargin=4pt,
frametitlebackgroundcolor=colorloc1]{ldBox}
\newenvironment{claimbox}{\begin{ldBox}\vspace{3 mm}
	} {\vspace{1.5 mm}\end{ldBox}}

\usepackage{fancyhdr}

\usepackage{fancyhdr}
\usepackage{sectsty}
\usepackage{titlesec}

\sectionfont{\color{colorloc5}\Large} 

\subsectionfont{\color{colorloc2}\large} 

\subsubsectionfont{\color{colorloc5}} 

\usepackage[linktocpage=true]{hyperref}
\hypersetup{colorlinks=true,linkcolor=colorloc5,citecolor=colorloc4,urlcolor=colorloc4}

\begin{document}

	%%%%%%%%%%%%%%%%%%%%%%%%%%%%%%%%%%%%%%%%%%%%%%%
	%%%%%%%%%%%%             				 %%%%%%%%%%%%%%%%%%%%%%
	%%%%%%%%%%%% 	  TITLEPAGE 	 %%%%%%%%%%%%%%%%%%%%%%
	%%%%%%%%%%%%             				%%%%%%%%%%%%%%%%%%%%%%
	%%%%%%%%%%%%%%%%%%%%%%%%%%%%%%%%%%%%%%%%%%%%%%%

	\baselineskip=14pt
	\parskip 3pt

	\vspace*{-1.5cm}
	%\begin{flushright}    % Publication numbers
	%	{\small 
		%		preprint?
		%	}
	%\end{flushright}
	
	\vspace{3cm}
	\begin{center}        % Main title

	{\color{colorloc5}\bf \huge Neural Network Learning and \\ Quantum Gravity}   
	\end{center}
	
	\vspace{0.5cm}
	\begin{center}        % Authors
		{\bf \large  Stefano Lanza}
	\end{center}
	
	%\vspace{0.1cm}
	\begin{center}  
		\emph{II. Institut f\"ur Theoretische Physik, Universit\"at Hamburg,\\
			Luruper Chaussee 149, 22607 Hamburg, Germany}
		\\[.3cm]
	\end{center}
	
	\vspace{3.5cm}

	\begin{abstract}
		\noindent The landscape of low-energy effective field theories stemming from string theory is too vast for a systematic exploration. However, the meadows of the string landscape may be fertile ground for the application of machine learning techniques. Employing neural network learning may allow for inferring novel, undiscovered properties that consistent theories in the landscape should possess, or checking conjectural statements about alleged characteristics thereof. The aim of this work is to describe to what extent the string landscape can be explored with neural network-based learning. Our analysis is motivated by recent studies that show that the string landscape is characterized by finiteness properties, emerging from its underlying tame, o-minimal structures. Indeed, employing these results, we illustrate that any low-energy effective theory of string theory is endowed with certain statistical learnability properties. Consequently, several learning problems therein formulated, including interpolations and multi-class classification problems, can be concretely addressed with machine learning, delivering results with sufficiently high accuracy.
	\end{abstract}

	\thispagestyle{empty}
	\clearpage
	
	\setcounter{page}{1}

%%%%%%%%%%%%%%%%%%%%%%%%%%%%%%%%%%%%%%%%%%%%%%%
%%%%%%%%%%%                 %%%%%%%%%%%%%%%%%%%
%%%%%%%%%%%  DOCUMENT BODY  %%%%%%%%%%%%%%%%%%%
%%%%%%%%%%%                 %%%%%%%%%%%%%%%%%%%
%%%%%%%%%%%%%%%%%%%%%%%%%%%%%%%%%%%%%%%%%%%%%%%

\newpage

\tableofcontents

\newpage

%%%%%%%%%%%%%%%%%%%%%%%%%%%%%%%%%%%%%%%%%%%%%%%
%%%%%%%%%%%                 										%%%%%%%%%%%%%%%%%%%
%%%%%%%%%%%  				INTRODUCTION			%%%%%%%%%%%%%%%%%%%
%%%%%%%%%%%                 										%%%%%%%%%%%%%%%%%%%
%%%%%%%%%%%%%%%%%%%%%%%%%%%%%%%%%%%%%%%%%%%%%%%

\section{Introduction}
\label{sec:Introduction}

Although string theory serves as a promising candidate for a theory of Quantum Gravity, its phenomenological implications are ambivalent and still unclear.
In fact, string theory leads to a gigantic number of consistent vacua, around which possibly different low-energy effective theories are defined.
Indeed, some estimations performed within F-theory compactifications  lead to roughly $10^{500}$ distinct vacua \cite{Ashok:2003gk,Denef:2004ze,Denef:2004dm}, but others yield even larger numbers, up to $10^{272,000}$ vacua \cite{Taylor:2015xtz}. 
Such a gargantuan number of vacua, and associated consistent effective field theories certainly cannot be examined in a systematic way.
It is thus hard to identify general properties and phenomenological implications of theories populating this large \emph{landscape} of consistent effective descriptions.

However, rather than studying effective theories of the string landscape systematically and exactly, a \emph{statistical} approach may be employed.
The emergence of big data sets in our contemporary society has propelled the birth and development of novel tools and techniques capable of extracting relevant information out of a large number of data. 
In recent years, machine learning techniques in general and neural networks in particular have acquired a critical role in retrieving knowledge from big data sets.
And the applications of such techniques are immense, ranging from finance to medicine, and new tools and methods are constantly developed, rendering theoretical and computational research in machine learning a vibrant area.
Indeed, machine learning techniques, eventually based on neural networks, may offer the right tools to explore the vast meadows of the string landscape.

In string theory contexts, machine learning techniques have been recently applied in order to understand properties of compactification manifolds and the associated moduli spaces \cite{Altman:2018zlc,Brodie:2019dfx,He:2020lbz,Anderson:2020hux,Constantin:2021for,Larfors:2021pbb,Abel:2023zwg,Berglund:2023ztk}; or to study some of the phenomenological features that string theory effective field theories are endowed with \cite{Ruehle:2017mzq,Carifio:2017bov,Cole:2019enn,Halverson:2020opj,Deen:2020dlf,Bies:2020gvf,Cole:2021nnt,Loges:2022mao,Halverson:2023ndu,Lanza:2023vee}, including, for instance, the search of models realizing inflation \cite{Dias:2018koa,Abel:2022nje}.

But the application of machine learning to string theory can go well beyond the aforementioned examples.
Exploiting supervised machine learning techniques one could predict properties that an effective theory in the landscape should be equipped with; or one could generically scan a large number of low-energy theories in order to check whether a given property is realized within some of the theories of the analyzed set.
Alternatively, the usage of unsupervised machine learning techniques may be helpful in order to uncover common patterns that occur in consistent effective field theories that would otherwise be too difficult to unearth via an analysis of the theories one by one.

Moreover, the usage of machine learning techniques can largely benefit the so-called \emph{Swampland program} as well.
The program, whose origins can be traced back to the seminal works \cite{Ooguri:2006in,Arkani-Hamed:2006emk}, aims at finding the common features that effective field theories stemming from string theory share; such features then deliver the prime criteria to discern whether an arbitrary low-energy effective field theory admits an ultraviolet completion within string theory, from a genuine bottom-up perspective.
Typically, the common features that string theory-originated effective theories allegedly share are proposed on the basis of some examples, and they are conjectural in nature.
Machine learning can then be beneficial for the program in a two-fold way: firstly, one can scan large data sets of distinct, consistent effective field theories in order to show whether a given conjectural feature is shared by these theories; secondly, by feeding a large number of consistent effective theories to a machine learning algorithm, new recurrent features could be revealed.

The common characteristic that all the aforementioned problems possess is the large computational complexity, which however renders these fit for an analysis via machine learning techniques. 
But can these Quantum Gravity problems be \emph{concretely} addressed using machine learning? 
Namely, are there machine learning-based algorithms that can \emph{learn} statistically significant answers to any of these problems?
In this work we will address these questions, equipped with two key tools.

Firstly, throughout Sections~\ref{sec:Deep_Learning_Definition} and~\ref{sec:VC_dimension}, we will state some of the main definitions and concepts of statistical learning theory.
In particular, we will introduce a mathematical definition of neural network learning that can be applied to a large variety of problems, from binary classification problems to general interpolations via real functions; additionally, we will present two quantities, the \emph{Vapnik-Chervonenkis dimension} and the \emph{fat-shattering dimension} that will help us quantify the learnability of a theory.
These notions, which have rarely appeared in theoretical physics literature, are pivotal to study the learnability of Quantum Gravity problems.

Secondly, the learnability of problems will be critically linked to the mathematical framework within which string theory-originated effective field theories are formulated, namely the one of \emph{o-minimal structures}, or of \emph{tame geometry} -- see, for instance, \cite{dries_1998} for a pedagogical introduction to the subject and for foundational references.
Indeed, it has been proposed that the couplings and the interactions that appear in consistent Quantum Gravity effective field theories are rather peculiar, for they must be \emph{definable} in a given o-minimal structure \cite{Grimm:2021vpn}, as we will review in Section~\ref{sec:o-minimal}.
The relevancy of such structures for string theory, and some of their phenomenological consequences have been explored in
\cite{Grimm:2021vpn,Bakker:2021uqw,Grimm:2022sbl,Grimm:2023lrf}, and they play an important role in more general quantum field theories as well \cite{Douglas:2022ynw,Douglas:2023fcg,Grimm:2023xqy}.

Armed with these tools from statistical learning theory and tame geometry, we will be able to state whether a problem formulated within a Quantum Gravity effective theory can be learned, for there exists an algorithm delivering a sufficiently precise answer.
In particular, we will show that the o-minimality of Quantum Gravity effective field theories \emph{implies} the learnability via neural networks of a large spectrum of problems, such as binary or multi-class classification problems or problems involving generic regression or interpolation.
In sum, our findings show that neural network learning within any effective description of Quantum Gravity is feasible, due to its underlying geometrical structure.

For concreteness, let us illustrate two examples of general problems within Quantum Gravity effective field theories that can be addressed via neural networks, and that can be learned, as we will show in the main text. 

\noindent\textcolor{colorloc3}{\textbf{Example I: Reconstructing the behavior of a gauge coupling.}} 
Consider a set of four-dimensional effective field theories that include a gauge field and a scalar field $\varphi$, that we regard as a modulus of the theory. 
The gauge field is associated with the gauge coupling $g(\varphi)$, which we assume to depend on the field $\varphi$.
The functional form of the gauge coupling in terms of the modulus $\varphi$ may be different for different effective field theories, since it depends on the details of the effective theory and its ultraviolet completion.
For instance, the specific form of $g(\varphi)$ may depend on the geometry of the manifold upon which string theory is compactified. 

\begin{figure}[thb]
	\centering
	\includegraphics[height=7cm]{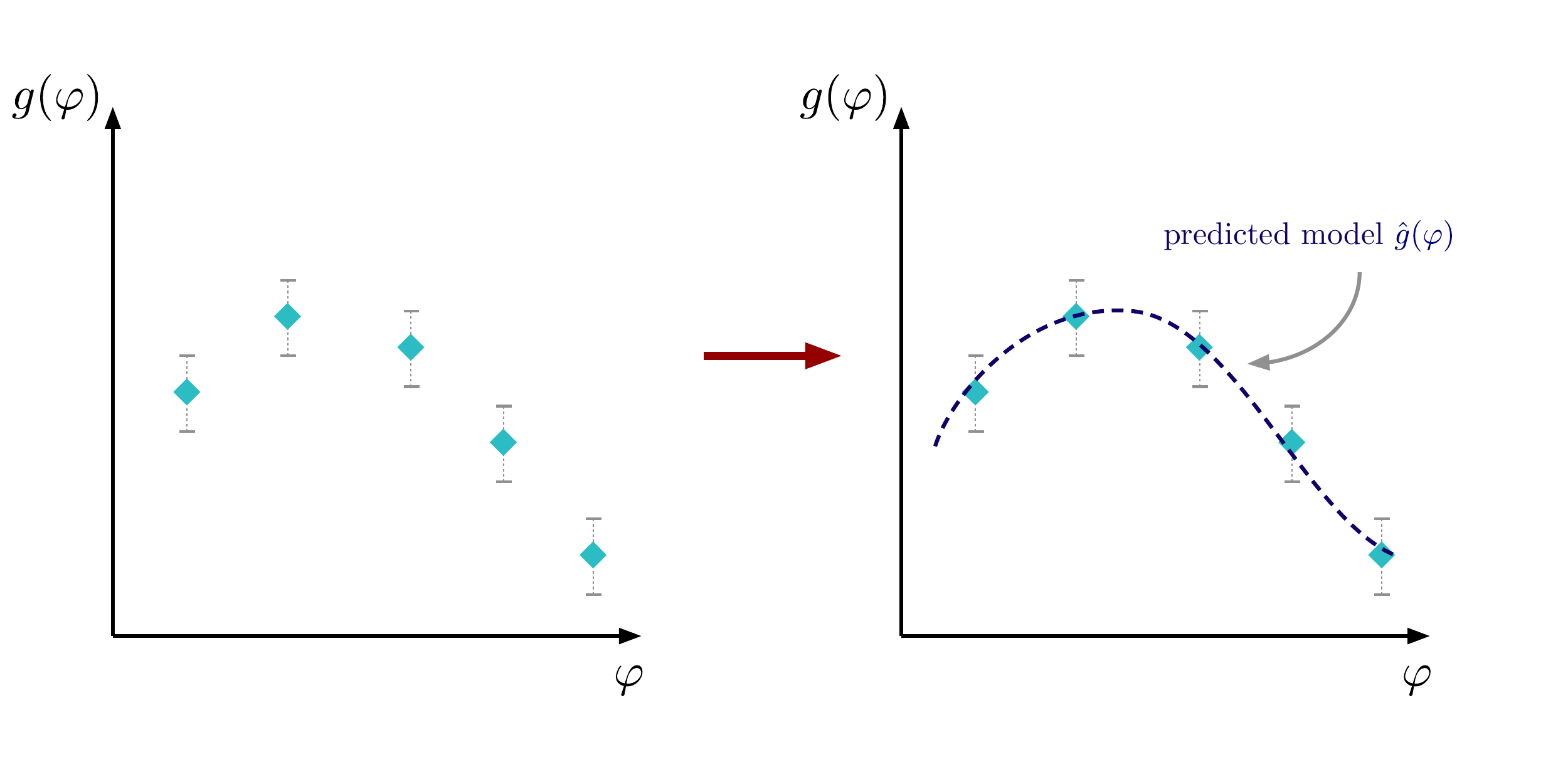}
	\caption{\footnotesize On the left is depicted a data set composed of the points $\{(\varphi_i, g_i)\}$. After applying a machine learning algorithm, one can infer the best model $\hat{g}(\varphi)$ that interpolates the data set, such as the one depicted on the right. \label{Fig:Intro_Example_1}}
\end{figure}

Assume that we do not know which to choose among the effective theories of the considered set. 
We could use the behavior of the gauge coupling as a `\emph{selection criterion}' for singling out a specific effective theory.
In particular, assume that we do know which values $g_i$, with $i = 1, \ldots, N_{\text{data}}$, the gauge coupling acquires for a discrete set of points $\varphi_i$ within the uni-dimensional moduli space spanned by the field $\varphi$ -- see Figure~\ref{Fig:Intro_Example_1} for a pictorial representation.
Ideally, the pairs $(\varphi_i, g_i)$ should be determined experimentally; alternatively, they can also be seen as theoretical constraints that are set in order, for the effective theory, to deliver a given phenomenology.

The `\emph{input}' data set, composed of all the points $(\varphi_i, g_i)$, can then be fed to a machine learning algorithm in order to single out which function $g(\varphi)$ better describes it. 
Namely, a machine learning algorithm may find the function $g(\varphi)$ for which it is `most likely' that the values $g(\varphi_i)$ are very close to the input values $g_i$.
Examples of machine learning techniques achieving this include general regression, or symbolic regression.
In turn, learning the shape of the function $g(\varphi)$ may tell crucial information about the geometrical data from which it descends, and about the effective theory in general.

It is worth stressing that, given the data set $(\varphi_i, g_i)$, one could always find a polynomial $g(\varphi)$ of degree $N_{\text{data}}$ passing through each of the points.
However, typically, this is not the goal: rather, one should look for a model that is sufficiently simple, with low complexity and such that it generalizes well to data points not included in the original data set $(\varphi_i, g_i)$.
Indeed, in Section~\ref{sec:o-minimal} we will see that Quantum Gravity prefers functions that are not `\emph{too general}', for they have to be defined in a given o-minimal structure.

\noindent\textcolor{colorloc3}{\textbf{Example II: Determining whether slow-roll inflation can be realized.}} Consider an effective field theory describing the dynamics of some real scalar fields $\varphi^\alpha$, with $\alpha = 1, \ldots, n$.
We assume the scalar fields to be subjected to the scalar potential $V(\varphi; c)$, with $c^{\mathsf{A}}$, $\mathsf{A} = 1, \ldots p$, being some parameters.
For instance, in effective theories obtained from a compactification of string theory, the parameters $c^{\mathsf{A}}$ may be internal geometric parameters, such as background fluxes.
We would like to inquire whether a given effective theory may accommodate slow-roll inflation.
To a first approximation this can be regarded as a \emph{boolean problem}: if for some parameters $c^{\mathsf{A}}$ the model can serve to realize slow-roll inflation, then the parameters $c^{\mathsf{A}}$ are associated with the label `\emph{true}'; otherwise, that choice of parameters cannot yield an effective theory accommodating slow-roll inflation and it will be labeled as `\emph{false}'. 
\begin{figure}[thb]
	\centering
	\includegraphics[height=7cm]{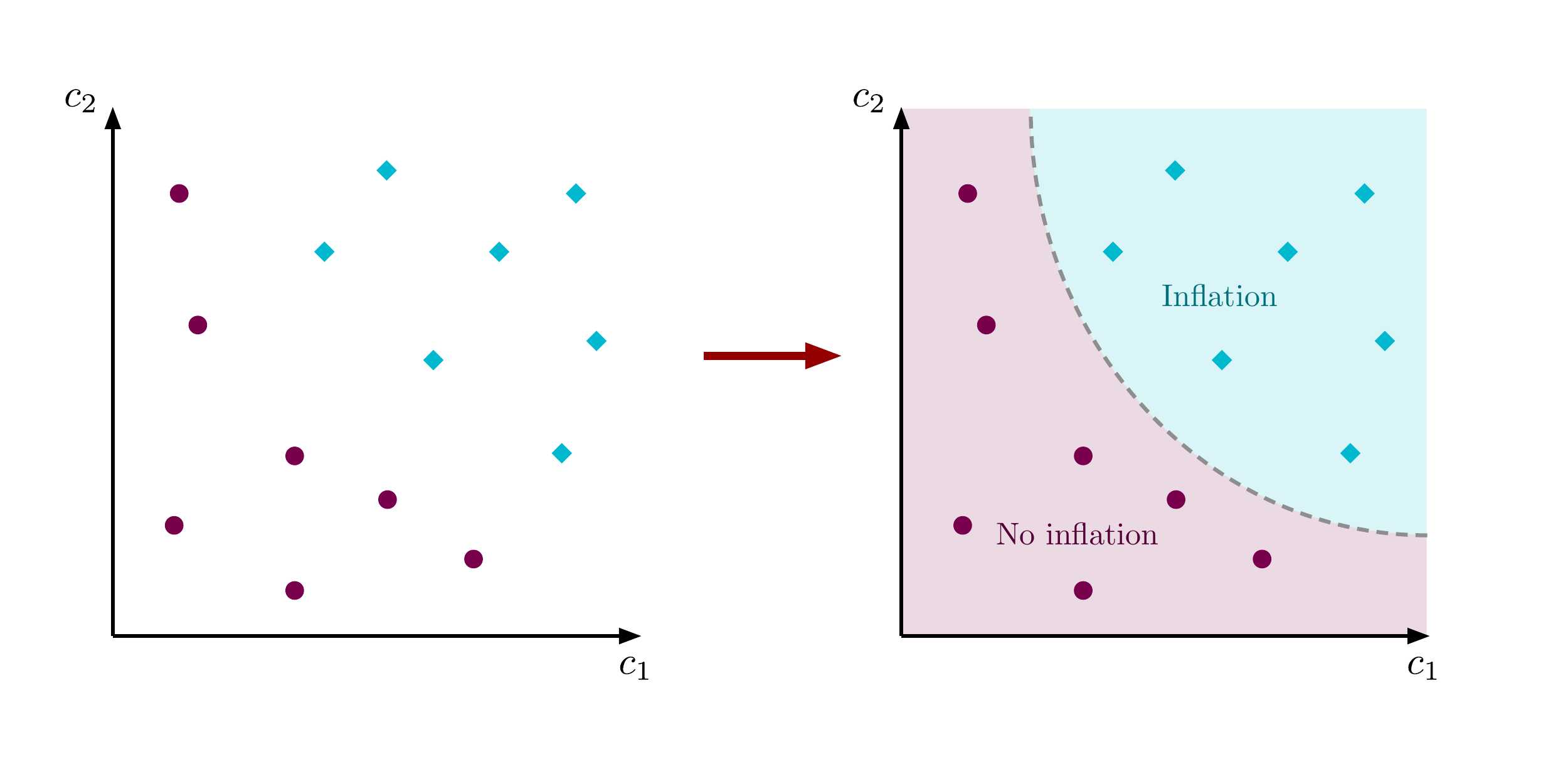}
	\caption{\footnotesize On the left is a data set composed of points in the parameter space spanned by $c_1$ and $c_2$:  the purple dots are those for which slow-roll inflation is not realized, while the light blue squares denote the points at which the slow-roll conditions are satisfied. On the right is an example of a boundary, which can be obtained via machine learning, that separates the regions of the parameter space in which inflation can or cannot be realized. \label{Fig:Intro_Example_2}}
\end{figure}
Assume that we do know whether slow-roll inflation is realized for a set of choices of parameters $\{ c^{\mathsf{A}}_i \}$, with $i = 1, \ldots, N_{\text{data}}$.
To each of the choices we associate a numerical label that carries such information: for instance, we could set $y = 1$ which stands for `\emph{true}', if the choice of parameters $c^{\mathsf{A}}_i$ leads to a model realizing slow-roll inflation, and $y = 0$ for `\emph{false}', if the model with the given parameters cannot realize slow-roll inflation.
The initial data set is then composed of the set of pairs $\{ (c^{\mathsf{A}}_i, y_i) \}$.
An example of such a data set is depicted on the left of Figure~\ref{Fig:Intro_Example_2}.

By feeding a data set of this form to a machine learning algorithm one could determine the regions in the parameter space spanned by $c^{\mathsf{A}}$ where it is more likely or less likely to realize slow-roll inflation.
For instance, applying machine learning techniques to the data set on the left of Figure~\ref{Fig:Intro_Example_2} could lead to the regions depicted on the right.

It is worth mentioning that a scan of possible inflationary models employing machine learning techniques has been performed in \cite{Abel:2022nje}.
In this work, we will address binary, boolean problems like this one from a more general perspective, and show that in Quantum Gravity problems like this one can be learned.

Let us summarize how this work is structured. 
The first three sections are dedicated to reviewing some definitions and concepts from statistical learning theory and tame topology that have rarely -- if not at all -- entered the high-energy physics literature.
Specifically, in Section~\ref{sec:Deep_Learning_Definition} we formally define the process of neural network learning, whose mathematical formulation will be pivotal for the sections that follow.
In Section~\ref{sec:VC_dimension} we introduce the concept of \emph{shattering} of a set, alongside with the related quantities, the \emph{Vapnik-Chervonenkis dimension} and the \emph{fat-shattering dimension}.
Then, Section~\ref{sec:o-minimal} serves as an overview of o-minimal structures and their role in Quantum Gravity effective field theories.
With the definitions and the tools introduced in Sections~\ref{sec:Deep_Learning_Definition},~\ref{sec:VC_dimension} and~\ref{sec:o-minimal}, in Section~\ref{sec:QG_Learnability} we present the main idea of this work: namely, we show that Quantum Gravity is naturally endowed with learning properties, and characterize how the learning is performed via theorems derived from statistical learning theory.
Finally, Appendix~\ref{sec:MT_o-min} collects additional information about the relation between o-minimal structures and the model theory independence property, upon which some conclusions of Section~\ref{sec:QG_Learnability} rely.

%%%%%%%%%%%%%%%%%%%%%%%%%%%%%%%%%%%%%%%%%%%%%%%
%%%%%%%%%%%                 										%%%%%%%%%%%%%%%%%%%
%%%%%%%%%%%  	DEFINING DEEP LEARNING		%%%%%%%%%%%%%%%%%%%
%%%%%%%%%%%                 										%%%%%%%%%%%%%%%%%%%
%%%%%%%%%%%%%%%%%%%%%%%%%%%%%%%%%%%%%%%%%%%%%%%

\section{Defining Learnability}
\label{sec:Deep_Learning_Definition}

The prime goal of this work is to assess whether some functions that appear in effective field theories stemming from Quantum Gravity, or functions related to those can be \emph{learned}. But what do we mean with `\emph{learned}'?
In this section we introduce a formal, mathematical definition of learning that is generically applicable to any neural network-based algorithm.
The definitions that we introduce here are mostly based on those given in \cite{anthony_bartlett_1999}, and are crucial for the discussion of the forthcoming sections.

\subsection{The learning framework}
\label{sec:Deep_Learning_Definition_Framework}

Firstly, we need to specify the general scope of the neural network learning that we consider and the framework within which the neural networks operate. 
In this work, we restrict our attention to \emph{supervised learning}: given some \emph{labeled} data, supervised algorithms learn the functions that relate the data to the respective labels, and are thus capable of predicting the label associated with new data.
Below we define how the data is structured for the learning process that we consider here, and how the functions that are learned are constructed.

The data set that is fed to the algorithm is composed of a set of vectors in $\mathbb{R}^n$, that we may denote as $x_i$ , with $i = 1, \ldots, N_{\text{data}}$, where $N_{\text{data}}$ is the number of data points initially available. 
Since the vectors $x_i$ are what is directly given to the algorithm, they constitute the \emph{inputs} of the neural network; moreover, their components -- which we denote as $x_{i,\alpha}$, so that $x_i = (x_{i,\alpha})^T$ -- take the name of \emph{features}.
Generically, we may assume that the \emph{input space} $X$, to which the vectors initially fed to the algorithm belong, is a compact subset of $\mathbb{R}^n$.
The subset $X$ should ideally comprise all the possible values that the inputs may take. 
However, the input vectors $x_i$ in our data set constitute only a subset of $X$, composed of isolated points and with finite cardinality. 

For example, in image processing, the input space may be given by the positions and colors of the pixels an image is composed of; or, in string phenomenological contexts, the input space may be a subset of the moduli space, with $x_i$ denoting the expectation values of some moduli in a given moduli space patch. 

In supervised learning problems, the initial data set also comes with \emph{labels}, which we denote as $y_i$.
Since, as we shall see below, the labels are what the algorithm aims to predict, the labels $y_i$ are said to belong to the \emph{output space} $Y$, the set of all the possible values that the labels may take.
In this work, we shall assume that the labels belong to a compact subset of $\mathbb{R}$, although more general choices are possible.
Whether $Y$ is a dense set depends on the specific problem at hand; relevant examples include:
\begin{description}
	\item[Binary classification:] the output space $Y$ is constituted solely by two points, $\{0\}$ and $\{1\}$; problems of this class answer boolean questions such as: \emph{does this image contain at least a person?} or \emph{does this effective theory allow for slow-roll inflation?}
	\item[Multi-class classification:] the output space $Y$ is constituted by a set of discrete points, which we may assume to be positive integers, and take $Y \subset \mathbb{N}$; examples of multi-class classification problems include: \emph{does this image contain a cat, a dog or neither of the two?} (with $Y = \{1, 2, 3\}$, encompassing the three possibilities) or \emph{which, among the Kaluza-Klein towers, first breaks down an effective field theory towards a given limit in the moduli space?} (where $Y = \{1, 2, \ldots, N_{\text{KK}}\}$, with $N_{\text{KK}}$ denoting the number of distinct Kaluza-Klein towers that could break down the effective theory in the given limit).
	\item[General regression or interpolation:] in this case, the output space is generically a compact subset of $\mathbb{R}$; problems belonging to this class aims at finding real-valued functions that map points in the input space to points in the output space; for instance, a regression problem aims at finding the general behavior of a gauge coupling defined over a subset of the moduli space.
\end{description}

In summary, the kind of data sets that we consider in this work are of the following form:
\begin{equation}
	\label{DL_data}
	\text{Data set} = \{  (x_i, y_i) \} \subset X \times Y \subset \mathbb{R}^{n+1}\,, \qquad i = 1, \ldots, N_{\text{data}}\,.
\end{equation}
In the following, we will oftentimes abbreviate the notation introducing the \emph{data vector} $z_i := (x_i, y_i)$, and the \emph{data space} $Z := X \times Y \subset \mathbb{R}^{n+1}$.

Given the data set \eqref{DL_data}, the key question that a supervised learning algorithm aims to address is the following:
\emph{consider a new input $x_{\rm{new}}$ that is not contained in the original input data set; what is the `most likely' label $\hat{y}_{\rm{new}}$ that can be associated with $x_{\rm{new}}$?}
For instance, for the image processing problem mentioned above, the algorithm aims at identifying whether a new image, not contained among the ones initially fed to the algorithm, contains a dog or a cat. 
Or, in the string phenomenological problem mentioned earlier, the algorithm would be able to predict the value of the gauge coupling on a new point in the moduli space.

In addressing this question, a key axiom is assumed to hold: namely, we shall assume that the data points in the set \eqref{DL_data} are picked from a \emph{true distribution} $P$, defined over $Z = X \times Y$, from which the data points in \eqref{DL_data} are randomly drawn.
Furthermore, in some geometrical problems, one can assume the stronger assumption that there exists a \emph{true function}
\begin{equation}
	\label{DL_true_function}
	f_{\text{true}}(x): \quad X \rightarrow Y\,,
\end{equation}
which models the data set \eqref{DL_data} and any other new point $(x_{\text{new}}, y_{\text{new}})$ \emph{without any mistake}. 
However, in general, the true function \eqref{DL_true_function} cannot be inferred; rather, one can find an \emph{estimated function} that is `sufficiently close' to the true one \eqref{DL_true_function}.
In order to achieve this, the algorithm first considers a \emph{family} of functions
\begin{equation}
	\label{DL_function}
	f(x,\omega): \quad X \times \Omega \rightarrow Y\,,
\end{equation}
Here, $\omega^{\mathsf{A}}$, with $\mathsf{A} = 1, \ldots, p$, denote a set of $p$ parameters, and $\Omega$ is the \emph{parameter space} to which $\omega^{\mathsf{A}}$ belong; we may assume that $\Omega$ is a compact subset of $\mathbb{R}^{p}$. 
In the following, we will call $\mathcal{F}$ the set of functions \eqref{DL_function}.
In more general non-geometrical problem, defining a `true function' as in \eqref{DL_true_function} may not make much sense, but one could still try to model the data via an estimated function of the form \eqref{DL_function} which, however, is not expected to model the data without making mistakes.

The choice of family of functions $\mathcal{F}$ that the algorithm ought to consider is tailored to the problem under consideration, and the dimension of $\Omega$ can be large.
As a simple example, if the family of functions is composed solely of lines in $\mathbb{R}^2$, then $\Omega$ is a two-dimensional subset, with a parameter $\omega^1$ given by the slope of the line, and another, $\omega^2$, representing the intercept.

After the class of functions \eqref{DL_function} is specified, the algorithm is \emph{trained} over the data set \eqref{DL_data}.
The purpose is to determine which is the \emph{best} function 
\begin{equation}
	\label{DL_function_best}
	f(x,\hat{\omega}): \quad X \rightarrow Y\,.
\end{equation}
that models the data set \eqref{DL_data}, mapping each of the inputs $x_i$ therein to a label $f(x_i, \hat{\omega})$ `as close as possible' to the given label $y_i$, for an appropriate choice of parameters $\hat{\omega}^{\mathsf{A}}$.
In this work we will not delve specifically into \emph{how} the process of finding the best parameters $\hat{\omega}^{\mathsf{A}}$ is performed, for this will be deferred to a future work.
However, we just mention that a typical way to get an estimation of the best parameters $\hat{\omega}^{\mathsf{A}}$ requires the definition of a `cost function' -- that depends on the parameters $\omega^{\mathsf{A}}$ and the data set points in \eqref{DL_data} -- and an optimization thereof that leads to the best parameters $\hat{\omega}^{\mathsf{A}}$.

Therefore, the result of this \emph{learning process} is to make the parameter space $\Omega$ collapse to the point $\hat{\omega}^{\mathsf{A}}$, representing the estimation for the parameters so that the estimated function \eqref{DL_function_best} is `sufficiently close' to the true function \eqref{DL_true_function}.
In turn, this allows for estimating the label $\hat{y}_{\text{new}}$ of any new input point $x_{\text{new}}$ as
\begin{equation}
	\label{DL_ynew}
	\hat{y}_{\text{new}} := f(x_{\text{new}} ,\hat{\omega})\,.
\end{equation}
In the upcoming section, we shall see how such a learning process leading to \eqref{DL_function_best} and estimations as \eqref{DL_ynew} can be formally defined.

\subsection{Defining Neural Network Learning}
\label{sec:Deep_Learning_Definition_Learning}

In the previous section we have stated the goal of a generic learning algorithm, but we have not specified what we mean when we say that `\emph{an algorithm learns a function}'.
Indeed, the aim of this section is to introduce a formal definition of \emph{learning algorithm}, which, on the one hand, refines the goal outlined in the previous section and, on the other hand, prescribes when the algorithm should stop the training.
Although such a definition can be particularized to the specific algorithm employed and to the optimization procedure upon which the algorithm is based, here, following~\cite{anthony_bartlett_1999}, we deliver a general definition of learning algorithm that is independent of how the algorithm specifically works.
For the sake of the clarity of the discussion, we first introduce a definition of learning algorithm that is fit for binary classification problems, which is easily extendable to multi-class classification problems, and later we show how learning algorithms can be more generically defined when dealing with general regressions or interpolations, for which the output is a generic real number.

\noindent\textcolor{colorloc3}{\textbf{Binary classification.}} 
Firstly, let us focus on a binary classification learning problem, for which the output space is composed of just two points, which we conventionally denote as $Y = \{0,1\}$, so that $Z = X \times \{0,1\}$.
Before coming to the very definition of learning for this class of problems, we preliminarily need a quantity capable of measuring `\emph{how well}' the algorithm is performing with respect to the training data set \eqref{DL_data}.
Thus, consider an estimated function $f(x, \hat{\omega})$ modeling the data set \eqref{DL_data}. 
We define the \emph{error of $f$ with respect to $P$} as
\begin{equation}
	\label{DL_er}
	\text{er}_P (f) = P \{ (x, y) \in Z : f(x, \hat{\omega}) \neq y \}\,.
\end{equation}
Namely, the error measures the probability (with respect to the true distribution $P$) that an arbitrary point $(x, y) \in Z$ is such that the predicted output $f(x, \hat{\omega})$ is different from the actual one $y$.
Hence, the closer the error is to one, the worse the estimation is.

Ideally, we would like to find a function $f_{\text{best}}$ within the set $\mathcal{F}$ of all the functions \eqref{DL_function} such that the error is the optimal one, defined as:
\begin{equation}
	\label{DL_opt}
	\text{opt}_P (\mathcal{F}) = \inf_{f \in \mathcal{F}} \text{er}_P (f) = \text{er}_P (f_{\text{best}})\,.
\end{equation}
We remark that, in general, the optimal error \eqref{DL_opt} is not zero: in fact, the set of functions $\mathcal{F}$ may not be large enough to include a function that correctly models all the points within the data set \eqref{DL_data}.
Moreover, as stressed earlier, for general neural network problems, the data set \eqref{DL_data} might not be described by a function at all (for instance, if the same $x_i$ is associated with multiple outputs, which may be possible due to the statistical nature of the problem).

Concretely, it might not be possible, or convenient to find a function within the family $\mathcal{F}$ with error equal to the optimal one \eqref{DL_opt}.
Indeed, if the family of functions $\mathcal{F}$ is too large, it could require too many iterations of the algorithm to find a function with the error \eqref{DL_opt}, thus rendering the algorithm inefficient.
Therefore, in practice, it is enough to find a function $f$ whose error is such that
\begin{equation}
	\label{DL_er_opt}
	\text{er}_P (f) < \text{opt}_P (\mathcal{F}) + \epsilon\,,
\end{equation}
for some $\epsilon > 0$, to be initialized by hand.

With the above definitions in mind, we are now in the position to define what we mean with `\emph{learning algorithm}':
\begin{importantbox}
	Consider a class of functions $\mathcal{F}$, with domain in the set $X$ and with values in the set $\{0,1\}$.
	A \emph{learning algorithm} is a function $\ell$:
	\begin{equation}
		\label{DL_L}
		\ell : \quad \bigcup\limits_{i = 1}^{\infty} Z^{(i)}  \to \mathcal{F}\,.
	\end{equation}
	The function $\ell$ has the property that, for any given $\epsilon \in ]0,1[$ and $\delta \in ]0,1[$, there exists an integer $m_0(\epsilon,\delta)$ such that
	\begin{itemize}[noitemsep,topsep=0pt]
		\item for $m \geq m_0(\epsilon,\delta)$,
		\item for any probability distribution $P$ defined over $Z = X \times \{0,1\}$\,,
	\end{itemize}
	given a training sample $z$ of length $m$
	\begin{equation}
		z = ((x_1, y_1), \ldots (x_m,y_m)) \in Z^{m}\,,
	\end{equation}
	then
	\begin{equation}
		\label{DL_L_err}
		\text{er}_P (\ell(z)) < \text{opt}_P(\mathcal{F}) + \epsilon
	\end{equation}
	holds with probability at least of $1 - \delta$, with $\delta$ being a \emph{confidence parameter}. 
	In other words, for $m \geq m_0(\epsilon,\delta)$, it has to hold that
	\begin{equation}
		\label{DL_L_P}
		P^{m} \{ \text{er}_P (\ell(z)) < \text{opt}_P(\mathcal{F}) + \epsilon \} \geq 1 - \delta \,.
	\end{equation}
	Additionally, if \eqref{DL_L_P} holds, we say that `\emph{$\mathcal{F}$ is learnable}'.
\end{importantbox}
Some comments about this definition of learning algorithm are in order.
Firstly, loosely speaking, the definition above can be understood as follows: an algorithm can be regarded as a `learning algorithm' if it can learn a function $f$, drawn out of the family $\mathcal{F}$, with arbitrary precision with respect to the optimal error \eqref{DL_opt}, provided that the training data set (whose dimension is measured by $m$) is sufficiently large.

Moreover, in general, we should expect that the algorithm can predict a function $f$ with greater accuracy for larger samples. 
In the definition above, the quantity $m_0(\epsilon,\delta)$, called `\emph{sufficient sample size}' delivers a possible lower bound on the dimension of the training set for which \eqref{DL_L_P} holds, but it may not be the lowest one.
Indeed, it is convenient to define the \emph{sample complexity} $m_{\ell}(\epsilon,\delta)$ of a learning algorithm $\ell$ such that \eqref{DL_L_P} holds, namely:
\begin{equation}
	\label{DL_mL}
	m_\mathscr{\ell}(\epsilon,\delta) = \min \{ m : \text{$m$ is a sufficient sample size for learning $\mathcal{F}$ by $\mathscr{\ell}$} \}\,.
\end{equation}

Clearly, the definition of sample complexity depends on the specific algorithm $\ell$.
However, one can also define the \emph{inherent sample complexity}:
\begin{equation}
	\label{DL_mF}
	m_{\mathcal{F}}(\epsilon,\delta) = \min\limits_{\mathscr{\ell}} m_\mathscr{\ell}(\epsilon,\delta)\,,
\end{equation}
which tells what is the absolute minimum sample size to learn a function within the family $\mathcal{F}$.

As we have stressed in Section~\ref{sec:Deep_Learning_Definition_Framework}, in general, the true probability distribution $P(x,y)$ is not known and the error \eqref{DL_er} may not be computable.
We can then try to \emph{estimate} the error \eqref{DL_er} starting from the data set \eqref{DL_data}.
For binary classification problems, a convenient estimation for the error \eqref{DL_er} is given by the sample error
\begin{equation}
	\label{DL_sample_er}
	\widehat{\text{er}}_z (f) = \frac{1}{N_{\text{data}}} | \{i: 1 \leq i \leq N_{\text{data}} \; \text{and}\; f(x_i, \hat{\omega}) \neq y_i \}|\,.
\end{equation}
Here, the brackets $| \cdot |$ denotes the \emph{cardinality} (namely, the number of elements) that the set within is composed of.
Thus, \eqref{DL_sample_er} is the averaged sum of all the points within the data set \eqref{DL_data} that are labeled incorrectly by the estimated function $f(x, \hat{\omega})$.
One could then scan across all the learning algorithms taking the training data set \eqref{DL_data} and with values in a given family of functions $\mathcal{F}$ such that the sample error is minimal, namely:
\begin{equation}
	\label{DL_sample_er_min}
	\widehat{\text{er}}_z (\ell(z)) = \min\limits_{f \in \mathcal{F}} \widehat{\text{er}}_z (f)\,,
\end{equation}
for any $z  \in Z^{N_{\text{data}}}$ and any $N_{\text{data}}$. 
We will call an algorithm achieving \eqref{DL_sample_er_min} a `\emph{sample error minimization algorithm}'.

Finally, we remark that, although here we have defined a learning algorithm for the case of binary classification problems, all the above definitions can be trivially extended to the cases where the output space has finite cardinality, and the above discussion can be applied to multi-class classification algorithms as well.

\noindent\textcolor{colorloc3}{\textbf{General regression or interpolation.}} The definition given above of learning algorithm can be generalized to general regression or interpolation problems with some modifications. 
Without loss of generality, we consider the output space to be the real, compact interval $[0,1]$; if the output space is a different compact interval, one can perform a transformation so that it reduces to the unit interval.

Being the output space a dense set, the definition of error \eqref{DL_er} is certainly not fit.
Rather, given a data point $z = (x,y)$, in this context we define the error committed by the model $f$ as
\begin{equation}
	\label{DL_er_real}
	\text{er}_P (f) = \mathbb{E}_P[(f(x) - y)^2] \,,
\end{equation}
where $\mathbb{E}_P$ denotes the expectation value of its argument, evaluated with respect to the true distribution $P$ out of which the data points are assumed to be drawn.
In particular, we assume that, although the probability distribution $P$ might have tails that go beyond the unit interval, it is sufficiently peaked around the ouput space.
Namely, we assume that there exists a constant $B \geq 1$ such that an output $y \in [1-B, B]$ with probability $1$.

It is worth mentioning that the argument of the expectation value in \eqref{DL_er_real}, $(f(x) - y)^2$, takes the name of `\emph{loss function}'.
Moreover, given \eqref{DL_er_real}, the optimal error is defined as the infimum of the error \eqref{DL_er_real} over the family of function $\mathcal{F}$, analogously to \eqref{DL_opt}.

Equipped with these definitions, one can define a learning algorithm for real-valued functions as follows.

\begin{importantbox}
	Consider a class of functions $\mathcal{F}$, with domain in the set $X$ and with values in the unit interval $[0,1]$.
	A \emph{learning algorithm} is a function $\ell$:
	\begin{equation}
		\label{DL_L_b}
		\ell: \quad \bigcup\limits_{i = 1}^{\infty} Z^{(i)}  \to \mathcal{F}\,.
	\end{equation}
	The function $\ell$ has the property that, for any given $\epsilon \in ]0,1[$, $\delta \in ]0,1[$ and $B \geq 1$, there exists an integer $m_0(\epsilon,\delta, B)$ such that
	\begin{itemize}[noitemsep,topsep=0pt]
		\item for $m \geq m_0(\epsilon,\delta,B)$,
		\item for any probability distribution $P$ defined over $Z = X \times [0,1]$ such that $y \in [1-B, B]$ with probability $1$ for any $y$,
	\end{itemize}
	given a training sample $z$ of length $m$
	\begin{equation}
		z = ((x_1, y_1), \ldots (x_m,y_m)) \in Z^{m}\,,
	\end{equation}
	then
	\begin{equation}
		\label{DL_L_err_b}
		\text{er}_P (\ell(z)) < \text{opt}_P(\mathcal{F}) + \epsilon
	\end{equation}
	holds with probability at least of $1 - \delta$, with $\delta$ being a \emph{confidence parameter}. 
	In other words, for $m \geq m_0(\epsilon,\delta,B)$, it has to hold that
	\begin{equation}
		\label{DL_L_P_b}
		P^{m} \{ \text{er}_P (\ell(z)) < \text{opt}_P(\mathcal{F}) + \epsilon \} \geq 1 - \delta \,.
	\end{equation}
	If \eqref{DL_L_P_b} holds, we say that `\emph{$\mathcal{F}$ is learnable}'.
\end{importantbox}

As for the binary classification case, one can define the \emph{sample complexity} $m_\ell(\epsilon,\delta,B)$ as the lowest $m_0(\epsilon,\delta)$ for which the above definition holds.
Moreover, for simplicity, in the following, we shall take $B = 1$, in such a way that the loss function also takes values in the unit interval.

As stressed above for the binary classification case, in general, the error \eqref{DL_er_real} is not computable, for the probability distribution $P$ is not known.
Hence, one can estimate the error \eqref{DL_er_real} starting from the training data set \eqref{DL_data} as
\begin{equation}
	\label{DL_er_real_est}
	\widehat{\text{er}}_z (f) = \frac{1}{N_{\text{data}}} \sum\limits_{i = 1}^{N_{\text{data}}} ( f(x_i; \hat{\omega}) - y_i )^2\,,
\end{equation}
which generalizes the estimation \eqref{DL_sample_er} to models with real outputs.
Then, in principle, one can look for a sample minimization algorithm obeying \eqref{DL_sample_er_min}.
However, for models with real output, the realization of \eqref{DL_sample_er_min} might be unfeasible, and realistically one can rather look for \emph{approximate sample minimization algorithms $\mathcal{A}(z,\epsilon)$} which satisfy the weaker condition
\begin{equation}
	\label{DL_approx_sem}
	\widehat{\text{er}}_z (\mathcal{A}(z,\epsilon)) < \inf_{f \in \mathcal{F}} \widehat{\text{er}}_z + \epsilon \,,
\end{equation}
for some fixed error $\epsilon$.

%%%%%%%%%%%%%%%%%%%%%%%%%%%%%%%%%%%%%%%%%%%%%%%
%%%%%%%%%%%                 										%%%%%%%%%%%%%%%%%%%
%%%%%%%%%%%  			THE VC DIMENSION		%%%%%%%%%%%%%%%%%%%
%%%%%%%%%%%                 										%%%%%%%%%%%%%%%%%%%
%%%%%%%%%%%%%%%%%%%%%%%%%%%%%%%%%%%%%%%%%%%%%%%

\section{Shattering dimensions}
\label{sec:VC_dimension}

In the last section we have defined how, starting with the training data set \eqref{DL_data}, a learning algorithm delivers a function, picked out of the family $\mathcal{F}$, that models the data \eqref{DL_data} within the error $\epsilon$ and with confidence parameter $\delta$.
In general, we should expect that increasing the size of the data set might render it more difficult to find a function within a given family $\mathcal{F}$, with the desired precision.
This might lead to an extension of the set of functions $\mathcal{F}$ in such a way that the predictivity of the model that the algorithm finds is ameliorated.
In this section we introduce the notion of `\emph{shattering}' of a subset, along with some quantities, namely the \emph{Vapnik–Chervonenkis dimension} and its generalization, the \emph{fat-shattering dimension}, that are capable of telling, given a data set \eqref{DL_data} and a set of functions $\mathcal{F}$, whether, within $\mathcal{F}$, there exists a function that can model the training data possibly with \emph{no mistakes}.

\subsection{The Vapnik–Chervonenkis dimension}
\label{sec:VC_dimension_Definition}

\begin{figure}[t]
	\centering
	\includegraphics[height=7cm]{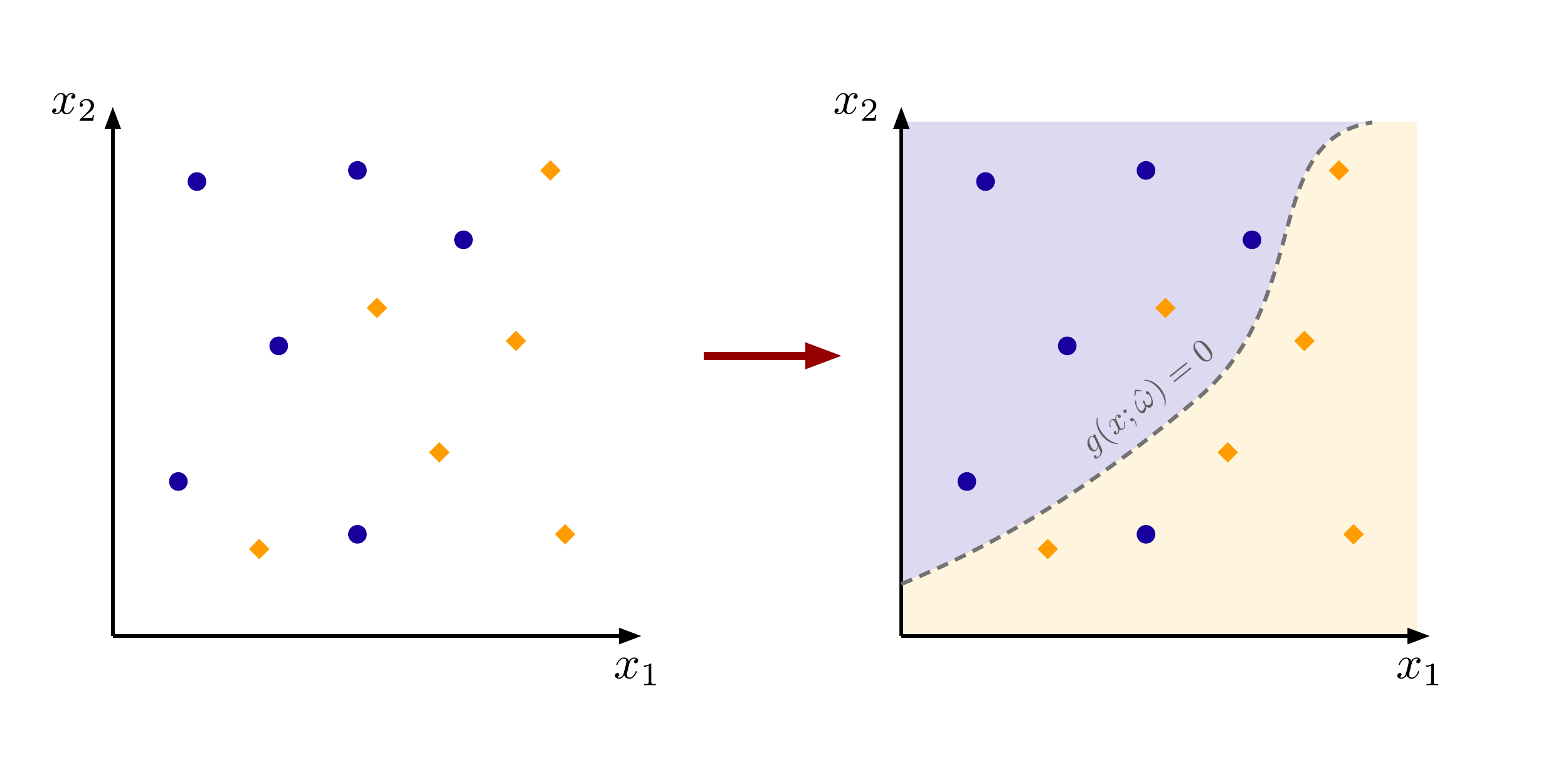}
	\caption{\footnotesize On the left is depicted a data set for a two-dimensional input space. The labels $\{0,1\}$ associated with a given point are here encoded in the shape and color of a point: a point with label $1$ is represented with a blue circle, and a point with label $0$ is represented as an orange square. On the right, the example of a function $g(x; \hat{\omega})$ that defines a partition of $\mathbb{R}^2$ via the locus $g(x; \hat{\omega}) = 0$ such that points above have predicted label $1$, and those below predicted value $0$. The function $g(x; \hat{\omega})$ depends on the estimated parameters and, using $g(x; \hat{\omega})$, one can define the actual model \eqref{DL_function_best}. Notice that, in general, the model does not predict the labels exactly. \label{Fig:shatter_gen_example}}
\end{figure}

Firstly, let us consider a simple binary classification problem, in which the output space is composed solely of two points, $Y = \{0,1\}$, and a family of functions $\mathcal{F}$ that map points of the input space to either $0$ or $1$. 
Let us focus on the restriction $\mathcal{F}|_{X_{\text{data}}}$ of these functions to the discrete set $X_{\text{data}}$ composed of the data points $\{x_i\}$.
The question that we wish to address is: \emph{within $\mathcal{F}|_{X_{\text{data}}}$ is there (at least) a function that predicts the correct output $y_i$ for every point $x_i$?}

It is convenient to introduce a different viewpoint on this problem that helps address such a question.
A function within $\mathcal{F}$ may be interpreted as a `boundary' that, in the full data input space $X$, separates the points $x_i$ into two sets, those with predicted output $0$ and those with predicted output $1$ -- see Figure~\ref{Fig:shatter_gen_example} for a pictorial representation.
In turn, restricting such a function to the discrete set $X_{\text{data}}$, the function selects a \emph{partition} of these points in two subsets, with labels associated with the predicted outputs $0$ and $1$.
Said differently, finding a function that models a given data set with binary outputs is equivalent to finding a function that selects an appropriate subset of $X_{\text{data}}$, within which the predicted output is $1$, with the complement set automatically associated with the label $0$.

Therefore, ideally, we would like to include, within the class of functions $\mathcal{F}$, those that are capable of predicting \emph{any} combination of the output, so that we are sure that, for any given data set, a model that appropriately classifies them can be found.
Specifically, recall that, for a set $S$ with cardinality $|S|$, one can construct $2^{|S|}$ subsets.
Thus, $\mathcal{F}$ should be such that \emph{all} these subsets can be modeled via some elements of $\mathcal{F}$.
The capability of a set of functions $\mathcal{F}$ to achieve this is captured by the \emph{Vapnik–Chervonenkis dimension}, whose definition relies on the concept of `\emph{shattering}' of a set.

Consider a family of sets $\mathcal{C}$ and a set $S$. We say that the set $S$ is \emph{shattered} by the sets of the family $\mathcal{C}$ if and only if the following family of sets 
\begin{equation}
	\label{VC_shatt}
	\mathcal{C} \cap S := \{ C \cap S  \, | \, C \in \mathcal{C} \}\,.
\end{equation}
 contains \emph{all} the subsets of $S$; namely, it has to hold that $| \mathcal{C} \cap S | = 2^{|S|}$.
Given a family of sets $\mathcal{C}$, the  \emph{Vapnik–Chervonenkis dimension of $\mathcal{C}$} is the cardinality of the largest set that $\mathcal{C}$ can shatter.
If any set can be shattered by the family of sets $\mathcal{C}$, we say that the Vapnik–Chervonenkis dimension of $\mathcal{C}$ is infinite.

These definitions can be most readily applied to the binary classification problem at hand by exploiting the considerations above, via the identifications
\begin{equation*}
	\begin{aligned}
		\text{Family of sets $\mathcal{C}$} \qquad &\leftrightarrow \qquad \text{Family of functions $\mathcal{F}$ (partitioning $X$ into subsets)} 
		\\
		\text{Set $S$ to shatter} \qquad &\leftrightarrow \qquad \text{Discrete set of data points $X_{\text{data}}$ to label}  
	\end{aligned}
\end{equation*}
As such, in the following, we will oftentimes call the Vapnik–Chervonenkis of the family of sets that the sets of functions $\mathcal{F}$ create with their partitions simply `the Vapnik–Chervonenkis of (the functions) $\mathcal{F}$'.
For the sake of concreteness, let us list some simple, though relevant examples of shattering induced by some families of functions.

\begin{figure}[t]
	\centering
	\includegraphics[width=13cm]{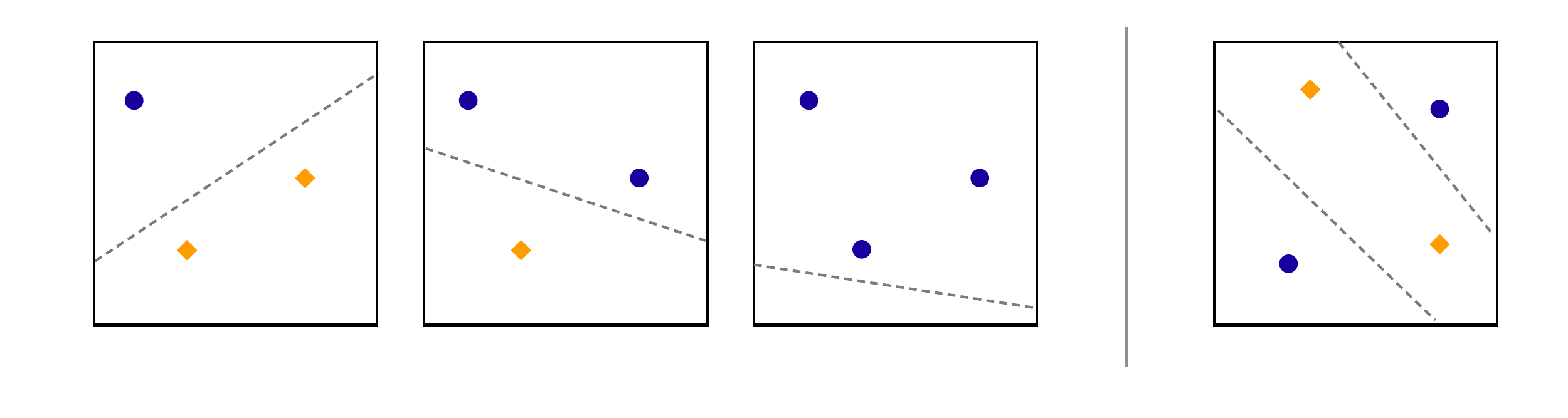}
	\caption{\footnotesize In $\mathbb{R}^2$, a line can always create a partition into two subsets such that three (non-collinear) data points with different labels (here, one label is represented by the data point being a blue circle, the other by the data point being an orange square) fall into the two distinct subsets. However, given a set of four points, a single line cannot always create a partition of $\mathbb{R}^2$ such that four points with different labels fall into the different partition components. \label{Fig:shatter_line_example}}
\end{figure}

\begin{description}[noitemsep,topsep=0pt]
	\item[Hyperplane classifier:] A simple example of class of functions $\mathcal{F}$, with domain in $\mathbb{R}^n$, achieving a binary classification is
	\begin{equation}
		\label{VC_sgnf}
		f(x;w,\theta) = \text{sgn}(w^\alpha x_\alpha + \theta )\,,
	\end{equation}
	where $\text{sgn}(x)$ is defined in such a way that $\text{sgn}(x) = 1$ if $x \geq 0$, and $\text{sgn}(x) = 0$ otherwise.
	In this case, the class of functions $\mathcal{F}$ is spanned by \eqref{VC_sgnf} varying the $n+1$ parameters $w^\alpha$ and $\theta$.

	For simplicity, consider, for instance, the simple case where the input space is a subset of $\mathbb{R}^2$, as depicted in Figure~\ref{Fig:shatter_line_example}.
	The locus $w^1 x_1 + w^2 x_2 + \theta = 0$ represents a line in the input space, and the function \eqref{VC_sgnf} delivers $1$ whenever a data point is above the line, and zero if it is below the line. 
	If our data set is composed of just one, two, or three (non-collinear) points, one can always find a line, by tuning the parameters $w^1$, $w^2$ and $\theta$ in such a way that \eqref{VC_sgnf} models the data exactly.
	However, if the data set is composed of four, or more points, one cannot always find a line that separates those into two subsets so that the predicted labels within them correspond to those of the data set. 
	Namely, lines cannot shatter sets of four points.
	Thus, in this example, the Vapnik–Chervonenkis dimension associated with the set of lines in $\mathbb{R}^2$ is three. 
	
	\item[Constant classifier:] As an extreme case, consider a function that is just a given constant. 
	This function is not fit for classifying any point, and thus the associated Vapnik–Chervonenkis dimension is zero.
	\item[Sine classifier:] Consider the case where the input space is the real line $\mathbb{R}$. The class of functions
	\begin{equation}
		\label{VC_sine}
		f(x; \theta) = \text{sgn}( \sin (\theta x) )\,,
	\end{equation}
	specified by the parameter $\theta$ can shatter arbitrarily large data sets. 
	As such, the Vapnik–Chervonenkis dimension of the sets determined by \eqref{VC_sine} is infinite.
\end{description}

\subsection{The fat-shattering dimension}
\label{sec:VC_fat-shattering}

The definition of shattering and of the Vapnik–Chervonenkis dimension given above rely on the fact that the output space is composed solely of two points, and cannot be easily extended to the case where the output space is composed of more than two points or where it is a dense set.
In the past decades, some extensions of the definition of Vapnik–Chervonenkis dimension have been proposed that analogously measure the possibility of correctly modeling the data with a given set of functions $\mathcal{F}$.
A prominent example is the Natarajan dimension \cite{Natarajan2004OnLS}, which extends the Vapnik–Chervonenkis dimension to the case where the output space is composed of a discrete set of points, as is for multi-class classification problems.
In this section we introduce the notion of \emph{fat-shattering}, employing which we define the \emph{fat-shattering dimension}, which applies to the more general regression cases, where the output space is a compact subset of the real line.

Consider a family of functions $\mathcal{F}$, defined over the input space $X \subset \mathbb{R}^n$, and with values in a compact subset of $\mathbb{R}$. 
As in Section~\ref{sec:Deep_Learning_Definition_Learning}, we assume that the functions in $\mathcal{F}$ take value in the unit interval $[0,1]$.
Analogously to the previous sections, it is also convenient to introduce $X_{\text{data}}$, the discrete subset composed of the input data points $x_i$.
Before introducing the very definition of fat-shattering dimension, we first define the $\gamma$-shattering. 

\begin{figure}[thb]
	\centering
	\includegraphics[height=9cm]{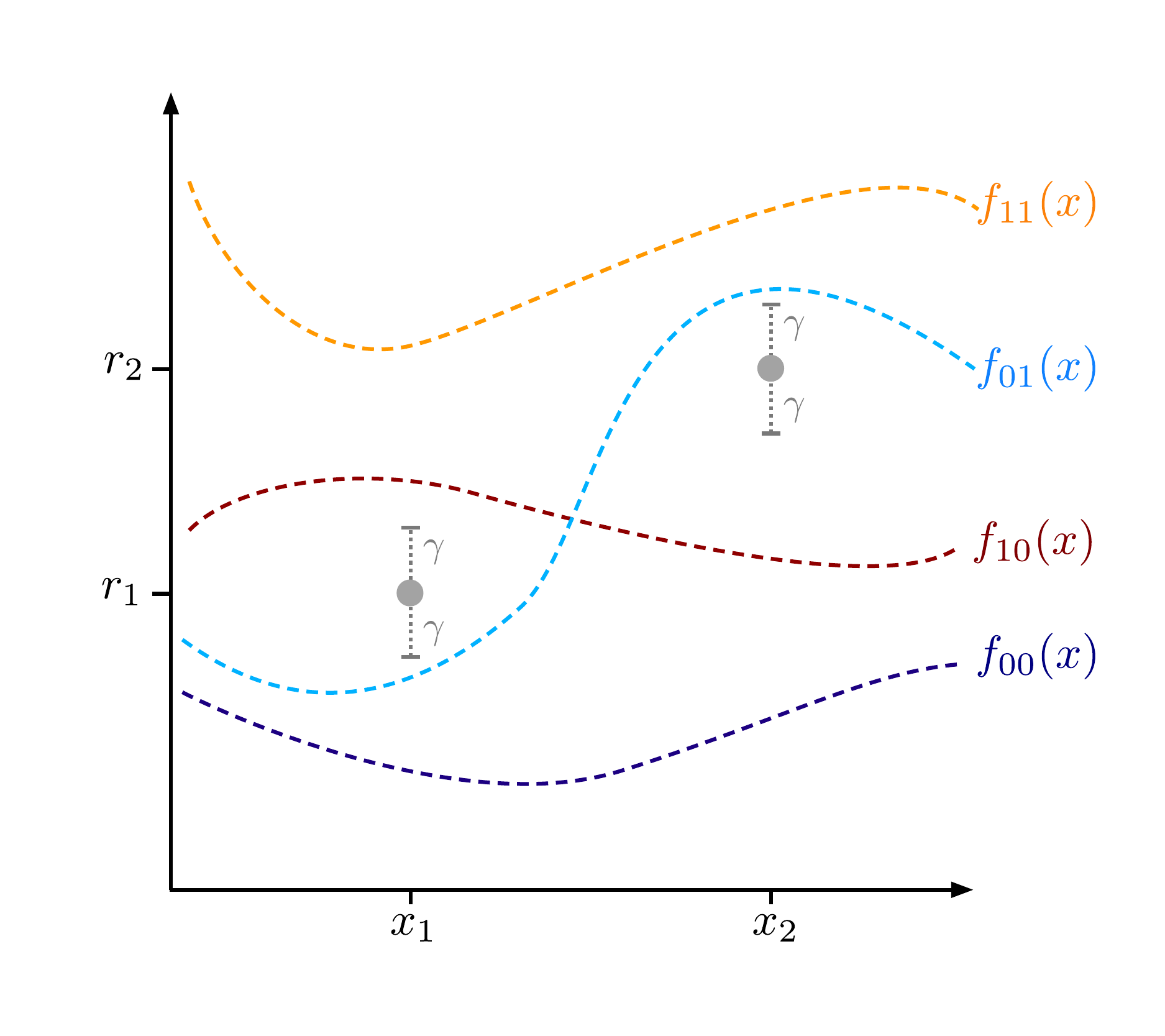}
	\caption{\footnotesize An example of $\gamma$-shattering for two data points in $\mathbb{R}$. Given $\gamma$, there exist two real numbers, $r_1$ and $r_2$, such that one can introduce four functions $f_{00}$, $f_{10}$, $f_{01}$ and $f_{11}$ obeying \eqref{VC_gamma-shatt_f}. Here, the subscripts of the functions denote the fictitious labels associated with $x_1$ and $x_2$. \label{Fig:gamma-shatter_example}}
\end{figure}

Assume to attribute some `fictitious' binary label $\{0,1\}$ to each of the data points $x_i$; we denote such a binary label, attributed to $x_i$, as $b_i$. 
Given $\gamma > 0$, we say that the input data set $X_{\text{data}}$ is \emph{$\gamma$-shattered} by the family of functions $\mathcal{F}$ if there exist some real numbers $r_1, \ldots, r_{N_{\text{data}}}$ such that, for any choice of $b_i$, there exists a function $f_b (x)$ obeying
\begin{equation}
	\label{VC_gamma-shatt_f}
	\begin{cases}
		f_b(x_i) \geq r_i + \gamma  &\text{if $b_i = 1$}\,,
		\\
		f_b(x_i) \leq r_i - \gamma &\text{if $b_i = 0$}\,.
	\end{cases}
\end{equation}
The point $r = (r_1, \ldots, r_{N_{\text{data}}})$ is said to `\emph{witness}' the shattering.
In Figure~\ref{Fig:gamma-shatter_example} is illustrated an example of such a $\gamma$-shattering for two points in $\mathbb{R}$.

Then, the \emph{fat-shattering dimension} $\text{fat}_\mathcal{F}(\gamma)$ is identified with the cardinality of the largest set that the family of functions $\mathcal{F}$ can $\gamma$-shatter.
It is worth stressing that, unlike the Vapnik–Chervonenkis dimension, the fat-shattering dimension $\text{fat}_\mathcal{F}(\gamma)$ does not solely depend on the family of functions $\mathcal{F}$, but also on the `width' of the shattering $\gamma$.
Indeed, we say that a family of functions has finite fat-shattering dimension if and only if $\text{fat}_\mathcal{F}(\gamma)$ is finite for every $\gamma > 0$.

%%%%%%%%%%%%%%%%%%%%%%%%%%%%%%%%%%%%%%%%%%%%%%%
%%%%%%%%%%%                 										%%%%%%%%%%%%%%%%%%%
%%%%%%%%  o-MINIMALITY and QUANTUM GRAVITY 		%%%%%%%%%%%%%%%%
%%%%%%%%%%%                 										%%%%%%%%%%%%%%%%%%%
%%%%%%%%%%%%%%%%%%%%%%%%%%%%%%%%%%%%%%%%%%%%%%%

\section{o-minimal structures in Quantum Gravity}
\label{sec:o-minimal}

As outlined in Section~\ref{sec:Deep_Learning_Definition}, a key step in setting up a learning algorithm consists in the specification of the family of functions \eqref{DL_function} among which the algorithm picks the best one modeling the input data \eqref{DL_data}.
Typically, the family of functions to be chosen is tailored to the specific problem under consideration.
In this section, we propose that, when dealing with a \emph{Quantum Gravity} learning problem, the functions that model the data set are restricted to be \emph{definable} in a given \emph{o-minimal structure}.
Below, based on \cite{dries_1998}, we briefly overview the definition of o-minimal structures and we recall what it means, for a function, to be definable in such a structure; then, we illustrate why these structures play a pivotal role in Quantum Gravity learning problems.

\subsection{o-minimal structures and definable functions}
\label{sec:o-minimal_Review}

Structures are some of the basic objects in model theory and can be regarded as concrete realization of languages.
However, in this work, we shall consider a special kind of `\emph{tame}' structures, the so-called \emph{o-minimal structures}, with `o' standing for `order'.
In the following, after introducing structures, we shall see how these may be rendered o-minimal, and how one can introduce functions that are `definable' in these structures. 
Here, we focus on the geometrical definitions of the objects, and we refer to Appendix~\ref{sec:MT_o-min_NIP} for model theory-oriented definitions.

\noindent\textcolor{colorloc3}{\textbf{Structures.}} 
Although they can be defined more generically, here we consider structures defined on \emph{sets}. A \emph{structure} $\mathcal{S}$ on a nonempty set $R$ -- denoted as $(\mathcal{S}, R)$ -- is a sequence $\mathcal{S} = (\mathcal{S}_n)_{n \in \mathbb{N}}$ such that, for each $n \geq 0$:
\begin{enumerate}[noitemsep,topsep=0pt]
	\item it is embedded with basic logical operations; namely, given two subsets $A, B \in \mathcal{S}_n$, $A \cup B \in \mathcal{S}_n$ and $R^n - A \in \mathcal{S}_n$;
	\item if $A \in \mathcal{S}_n$, then $R \times A \in \mathcal{S}_{n+1}$ and $A \times R \in \mathcal{S}_{n+1}$;
	\item the loci $\{(v_1, \ldots, v_n) \in \mathbb{R}^n:\; v_1 = v_n  \} \in \mathcal{S}_{n}$;
	\item given the projection map on first $n$ coordinates $\pi(A): R^{n+1} \to R^n$, if $A \in \mathcal{S}_{n+1}$, then $\pi(A) \in \mathcal{S}_{n}$.
\end{enumerate}

Notice that the first property implies that also $R^n \in \mathcal{S}_n$, $\varnothing \in \mathcal{S}_n$ and $A \cap B \in \mathcal{S}_n$.

\noindent\textcolor{colorloc3}{\textbf{Functions and structures.}} 
Having defined structures on sets, one can introduce functions that map from one set within the structure to another set within it. 
In turn, functions define sets via their graphs, and it is natural to introduce the following definition. 
Consider a structure $(\mathcal{S}, R)$, and a function $f: A \to R^n$, with $A \subset R^m$. 
If the graph of the function $f$, $\Gamma(f) \subset R^m \times R^n$, is contained within $\mathcal{S}_{m+n}$, then we say that `the function $f$ is definable within the structure $\mathcal{S}$'. 

Definable functions are quite peculiar, and they are equipped with several properties. 
Consider a subset $A \subset R^m$, and a map $f: A \to R^n$ that is definable within the structure $\mathcal{S}$, then:
\begin{itemize}[noitemsep,topsep=0pt]
	\item $A \in \mathcal{S}_m$;
	\item if $S \subset A$, with $S \in \mathcal{S}_m$, then also $f(S) \in \mathcal{S}$ and the restriction $f|_S \in \mathcal{S}$;
	\item if $B \in \mathcal{S}_n$, then the inverse $f^{-1}(B) \in \mathcal{S}_m$;
	\item if $f$ is injective, then its inverse $f^{-1}$ belongs to $\mathcal{S}$;
	\item given a second function $g: B \to R^p$, such that $f(A) \subseteq B$, then the composite function $g \circ f: S \to R^p$ belongs to the structure $\mathcal{S}$.
\end{itemize}

\noindent\textcolor{colorloc3}{\textbf{o-minimal structures.}} Enforcing some `\emph{tame}' properties on the structures defined above, one can introduce \emph{o-minimal structures}, defined inductively as follows.
Consider a \emph{linearly order} dense set $(R, <)$, with the symbol `$<$' being the binary relation specifying the order.
We assume $R$ to be without endpoints (namely, one cannot single out either the largest or the smallest element within $R$).
A structure $(\mathcal{S}, R)$ is said to be \emph{o-minimal} if the following, additional properties hold
\begin{enumerate}[noitemsep,topsep=0pt]
	\item the sets in $\mathcal{S}_1$ are \emph{finite} unions of points or intervals;
	\item the set $\{(v_1, v_2) \in R^2:\; v_1 < v_2 \} \in \mathcal{S}_2$. 
\end{enumerate}

As is clear from the definition, o-minimal structures are distinguished from general structures due to two key aspects:
\begin{description}[noitemsep,topsep=0pt]
	\item[Ordering:] We need that the set $R$ over which the structure is defined to be \emph{linearly ordered}. Recall that, loosely speaking, `\emph{linear order}' means that, given any two elements within the structure, these can be always compared.\footnote{Formally, introducing the binary relation `$\leq$', we say that a set $R$ is equipped with linear order if, given any $a, b, c \in R$, then: $a \leq a$; if $a \leq b$ and $b \leq a$, then $a = b$; ; if $a \leq b$ and $b \leq c$, then $a \leq c$; either $a \leq b$ or $b \leq a$. By `modding out' the equality via negations, one can in turn define the strict inequality symbol `$<$'.} Indeed, this ordering property gives the name to this kind of structures;
	\item[Finiteness:] Enforcing the finiteness requirement on $\mathcal{S}_1$ reflects on the other elements of the sequence: any $\mathcal{S}_n$ can only be composed of a \emph{finite} number of components. 
\end{description}

The definitions delivered above might appear a little abstract.
In order to render the discussion more concrete, in the following, restricting to the case where the set $R$ is the real line $\mathbb{R}$, we illustrate how to recognize functions that are definable in o-minimal structures and we deliver some examples of o-minimal structures

\noindent\textcolor{colorloc3}{\textbf{Functions definable in o-minimal structures.}} In order to recognize whether a function can be defined in a given o-minimal structure, the following \emph{Tame monotonicity theorem} is helpful:
\begin{importantbox}
	Given a definable function $f: A \to \mathbb{R}$, with $A \subset \mathbb{R}^m$, there is a finite partition of $A$ into regular cells such that, on each of these cells, $f$ is either strictly increasing, strictly decreasing or constant.
\end{importantbox}
Here, regular cells are defined as follows. 
Consider a sequence $(i_1, \ldots, i_m)$ of zeros and ones. 
An $(i_1, \ldots, i_m)$-cell is a subset of $\mathbb{R}^m$ defined inductively such that:
\begin{enumerate}[noitemsep,topsep=0pt]
	\item a $(0)$-cell is a point over the real line $\mathbb{R}$, while a $(1)$-cell is an open interval of $\mathbb{R}$ (which may be also of the form $]-\infty, a[$ or $]b, +\infty[$, for some $a, b \in \mathbb{R}$);
	\item given the $(i_1, \ldots, i_n)$-cells, a cell $(i_1, \ldots, i_n,0)$ is the graph $\Gamma(f)$ of a function $f: C \to \mathbb{R}$, with $C$ an $(i_1, \ldots, i_n)$-cell, and a $(i_1, \ldots, i_n,1)$-cell is a set delimited by the graph of two functions; namely, given $f, g: C \to \mathbb{R}$, a $(i_1, \ldots, i_n,1)$-cell is given by
	\begin{equation}
		(f,g) := \{ (x, r) \in C \times \mathbb{R} : \; f(x) < r < g(x) \}\,.
	\end{equation}
\end{enumerate}
Furthermore, a cell $C \subset \mathbb{R}^m$ is called `regular' if, given two points $x, y \in C$ that differ only in the $i$-th coordinate and such that $x^i < y^i$, then any $z^i$ such that $x^i < z^i < y^i$, we have $z^i \in C$. 

The Tame monotonicity theorem has two key consequences. 
Firstly, definable functions have regular `tails': for large enough coordinates, definable functions can only be either strictly increasing or strictly decreasing or constant, and no other possibility is allowed.
Moreover, definable functions need to have only a finite number of loci where the function has critical values.
In other words, it is not possible that a definable function is equipped with an infinite number of critical loci, otherwise it would violate the Tame monotonicity theorem.

\begin{figure}[thb]
	\centering
	\includegraphics[width=7cm]{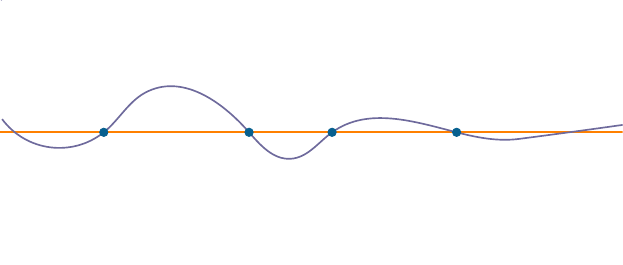} \qquad \includegraphics[width=7cm]{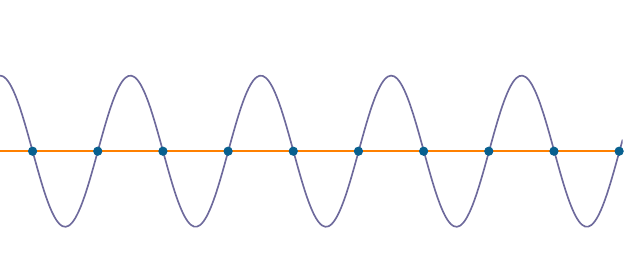}
	\caption{On the left, an example of a function that can be definable in a structure and, on the right, an example of a function that is not definable. \label{Fig:o-min_examples}}
\end{figure}

Thus, employing the Tame monotonicity theorem, one can most readily tell whether a function is not definable within a structure.
In Figure~\ref{Fig:o-min_examples} are depicted two examples: in the plot on the left is a function that has only a finite number of critical points, and only a finite number of intervals where it is either strictly decreasing or increasing; therefore, the function on the left in Figure~\ref{Fig:o-min_examples} can be definable in a given o-minimal structure. 
Instead, the plot on the right of Figure~\ref{Fig:o-min_examples} depicts a periodic function that displays an infinite number of critical points and, as such, is strictly increasing or decreasing on an \emph{infinite} number of intervals; hence the function on the right in Figure~\ref{Fig:o-min_examples} cannot be definable in an o-minimal structure.

\noindent\textcolor{colorloc3}{\textbf{Examples of o-minimal structures.}} Above we have seen how, given an o-minimal structure, one can introduce definable functions. 
However, it is quite hard to determine which set of functions one can include in an o-minimal structure while maintaining the o-minimality property.
Specifically, let us consider an o-minimal structure $\mathcal{S}$ defined over the real line, and let us include a set of real-valued functions $\mathcal{A}$, with domains in $\mathbb{R}^m$. 
We denote such an object as $\mathbb{R}_{\mathcal{A}}$. Whether $\mathbb{R}_{\mathcal{A}}$ is an o-minimal structure is a highly non-trivial problem, and the allowed choices of $\mathcal{A}$ that render $\mathbb{R}_{\mathcal{A}}$ an o-minimal structure is a main topic of current research in o-minimal structures.
  
In the past decades, the following three, key o-minimal structures have been identified:
\begin{description}[noitemsep,topsep=0pt]
	\item[$\mathbb{R}_{\text{alg}}$:] this structure is composed of semi-algebraic sets of the form:
	\begin{equation}
		\label{o-min_Ralg}
		\{ x \in \mathbb{R}^n: \; P_1(x) = \ldots = P_k(x) = 0\,, Q_1(x) > 0 , \ldots, Q_l(x) >0 \}\,,
	\end{equation}
	where $P_1(x), \ldots, P_k(x)$ and $Q_1(x), \ldots, Q_l(x)$ are polynomials in the $x^i$ variables. 
	As such, this structure includes all the functions that can be understood as loci of the form \eqref{o-min_Ralg}, such as polynomials, or functions such as $y = x^{\frac1n}$, for some integer $n$.
	\item[$\mathbb{R}_{\text{exp}}$:] this structure is an extension of the $\mathbb{R}_{\text{alg}}$ so as to include the exponential function and the logarithm;
	\item[$\mathbb{R}_{\text{an,exp}}$:] this structure extends $\mathbb{R}_{\text{exp}}$ so as to include \emph{restricted analytical functions}, and their inverses. 
	Recall that a restricted analytical function $f$ over a cube $C(a, \varepsilon) = \{ x \in \mathbb{R}^n: \; \forall\, i\; |x_i - a_i| \leq \varepsilon \}$ is a function whose restriction $f|_{C(a, \varepsilon)}$ is analytic and it can be analytically extended elsewhere.
\end{description}

\subsection{Tameness and Quantum Gravity Learning}
\label{sec:o-minimal_Quantum_Gravity}

The `tame' o-minimal structures introduced in the previous section stem from model theory, and they have been pivotal in defining a new field of mathematical research called \emph{tame topology}.
However, they did not find applications in Physics until recently: the newly proposed \emph{Tameness Conjecture} \cite{Grimm:2021vpn} asserts the prominence of o-minimal structures in effective field theories that admit an ultraviolet completion within a Quantum Gravity theory, such as string theory.

In order to illustrate the statement and implications of the Tameness Conjecture, let us consider a general $D$-dimensional effective field theory valid up to a given energy cutoff $\Lambda_{\text{\tiny{EFT}}}$, that we assume to be smaller than the Planck mass $M_{\rm P}$. 
For example, such an effective theory may include the following bosonic terms:
\begin{equation}
	\label{o-min_Action}
	\begin{aligned}
		S^{(D)} &= \int \Big( \frac{1}2 M^{D-2}_{\rm P} R \star 1 - \frac12 M^{D-2}_{\rm P} G_{ab}(\varphi,\lambda) {\rm d} \varphi^a \wedge \star {\rm d} \varphi^b 
		\\
		&\qquad\qquad - \frac12 M^{D-2(p_\mathcal{I}+1)}_{\rm P} f_{\mathcal{I}\mathcal{J}}(\varphi,\lambda)   F^\mathcal{I}_{p_{\mathcal{I}}+1} \wedge * F^\mathcal{J}_{p_{\mathcal{J}}+1} - V(\varphi,\lambda) \star 1 + \ldots \Big)\,.
	\end{aligned}
\end{equation}
where $R$ denotes the $D$-dimensional Ricci scalar and `$\star$' is the $D$-dimensional Hodge-duality operator. The action \eqref{o-min_Action} contains the following ingredients:
\begin{itemize}[noitemsep,topsep=0pt]
	\item a set of real parameters $\lambda^\kappa$, $\kappa = 1\, \ldots, k$, defined over a subset $\mathcal{P}$ of $\mathbb{R}^k$; in top-down constructions, the parameters $\lambda^\kappa$ may collect internal geometrical data such as background fluxes;
	\item a set of real scalar fields $\varphi^{a}$, $a = 1, \ldots, N$; they parametrize a local chart of the (unconstrained) $N$-dimensional moduli space $\mathcal{M}_\lambda$. In general, we here understand the moduli space as a fibration over the parameter space $\mathcal{P}$. Thus, in the local chart parametrized by the scalar fields $\varphi^{a}$ one can define the field space metric $G_{ab}(\varphi,\lambda)$ that depends on both the fields $\varphi^a$ and the parameters $\lambda^\kappa$. Additionally, we have assumed the presence of a scalar potential $V(\varphi,\lambda)$ that may introduce constraints on the moduli space $\mathcal{M}_\lambda$;
	\item a set of abelian $p$-form gauge fields $A^\mathcal{I}_{p_{\mathcal{I}}}$, $\mathcal{I} = 1, \ldots, M$, whose field strengths are defined as $F^\mathcal{I}_{p_{\mathcal{I}}+1} = {\rm d}A^\mathcal{I}_{p_{\mathcal{I}}}$. The gauge kinetic function $f_{\mathcal{I}\mathcal{J}}(\varphi,\lambda)$ that encodes the gauge couplings is generically defined over the fibration $\mathcal{M}_\lambda$.
\end{itemize}

The Tameness Conjecture proposed in \cite{Grimm:2021vpn} restrains the moduli and parameter spaces, as well as the functional form of any coupling appearing in the action \eqref{o-min_Action} in that it asserts that:
\begin{itemize}[noitemsep,topsep=0pt]
	\item the parameter space $\mathcal{P}$ and the fibered moduli space $\mathcal{M}_\lambda$ need to be definable in a given o-minimal structure;
	\item any coupling $g(\varphi,\lambda)$ that appears in the effective action \eqref{o-min_Action} must to be definable in the given o-minimal structure.
\end{itemize}

Concretely, consider a generic coupling $g(\varphi,\lambda)$ defined over $\mathcal{M}_\lambda$. In the effective theory described by the action \eqref{o-min_Action} such a coupling can be, for instance, any of the elements of the metric $G_{ab}(\varphi,\lambda)$, a gauge coupling defined through the gauge kinetic function $f_{\mathcal{I}\mathcal{J}}(\varphi,\lambda)$ or the scalar potential $V(\varphi,\lambda)$. 
The definability of the coupling $g(\varphi,\lambda)$ implies that $g(\varphi,\lambda)$ can be written as originating from the following locus:
\begin{equation}
	\label{o-min_Loci}
	\begin{aligned}
		\exists\; x_1, \ldots, x_l: \qquad &P_i(\varphi,\lambda,x,g, f_1, \ldots, f_m )=0\,,
		\\
		&Q_j(\varphi,\lambda,x,g, f_1, \ldots, f_m)>0\,,
	\end{aligned}
\end{equation}
or as unions, intersections and complements of loci of the form \eqref{o-min_Loci}.
Here $P_i$ and $Q_j$ are polynomials in their arguments. 
Besides the local moduli $\varphi^a$ and the parameters $\lambda^\kappa$, here we have 
the `auxiliary variables' $x_1, \ldots, x_l$ that may be needed to construct the locus \eqref{o-min_Loci} within the o-minimal structure.
Indeed, only in structures that admit quantifier elimination one can rewrite the locus \eqref{o-min_Loci} without the usage of the auxiliary variables $x_1, \ldots, x_l$ and the quantifier `$\exists$'. Moreover, $f_1, \ldots, f_m$ denote functions of the other arguments, whose form depends on the specific o-minimal structure under consideration.
For instance, in an $\mathbb{R}_{\text{exp}}$ o-minimal structure, the functions  $f_1, \ldots, f_m$ may be exponential or logarithms of the other variables.

However, it is worth remarking that, although the discussion above focuses on the action \eqref{o-min_Action}, the statements of the Tameness Conjecture hold for more general actions that include -- for instance -- non-abelian gauge fields, fermionic couplings or higher-derivative interactions. 
Additionally, the Tameness Conjecture does not rely on whether the effective theory preserves any supercharge and the action \eqref{o-min_Action} may well be non-supersymmetric.

Furthermore, in some instances, one can \emph{prove} the assertions of the Tameness Conjecture.
Indeed, exploiting the results of \cite{BKT}, one can show that the complex structure moduli space of effective theories obtained from F-theory compactified over Calabi-Yau four-folds lies within the $\mathbb{R}_{\text{an,exp}}$ o-minimal structure \cite{Grimm:2021vpn}. A similar conclusion holds for the complex structure moduli space of effective theories obtained from compactification of Type IIB string theory over Calabi-Yau three-folds \cite{Grimm:2022sbl}.
In particular, in the latter case, it can be shown that the couplings related to the vector multiplet sector can be defined in the $\mathbb{R}_{\text{an,exp}}$ o-minimal structure.

Within such classes of effective field theories, relevant examples of couplings that can appear and that are definable in $\mathbb{R}_{\rm an,exp}$ include:
\begin{itemize}[noitemsep,topsep=0pt]
	\item the ring $\mathcal{O}[\varphi]$ of polynomials in the real scalar fields $\varphi^a$ with coefficients belonging to $\mathcal{O}$, which is the set of real restricted analytic functions in $(\varphi,\lambda)$; namely, a coupling $g \in \mathcal{O}[\varphi]$ exhibits a general expansion of the form
	\begin{equation} 
		\label{Defin_Ex1}
		g(\varphi,\lambda) = \sum_{ \mathbf{m}}\rho_{\mathbf{m}}(\varphi,\lambda)   (\varphi^1)^{m_1} \cdots (\varphi^m)^{m_n}\, , 
	\end{equation}
	with ${\bf m} = \{m_1, \ldots, m_n\} \in \mathbb{Z}^n$ and $\rho_{\mathbf{m}}(\varphi,\lambda)$ restricted analytic functions in $(\varphi,\lambda)$. Notice that also couplings of the form \eqref{Defin_Ex1} with \emph{real} exponents $m_1, \ldots, m_n$ are tame;
	\item couplings that contain real exponentials of the real scalar fields $\varphi^a$ as in
	\begin{equation} 
		\label{Defin_Ex2}
		g(\varphi,\lambda) = \sum_{ m}\rho_{m}(\varphi,\lambda) \exp\left[ P_m (\varphi) \right]\, , 
	\end{equation}
	with $\rho_{m}(\varphi,\lambda)$ restricted analytic functions in $(\varphi,\lambda)$ and $P_m$ polynomials in $\varphi$. Moreover, also couplings with iterated exponentials, such as
	\begin{equation} 
		\label{Defin_Ex2b}
		g(\varphi,\lambda) = \sum_{ m}\rho_{m}(\varphi,\lambda) \exp\left[\exp \left[ \ldots \exp\left[ P_m (\varphi) \right]\right]\right]\, ,
	\end{equation}
	are definable in $\mathbb{R}_{\text{an,exp}}$;
	\item couplings that contain periodic functions, such as
	\begin{equation} 
		\label{Defin_Ex3}
		g(\varphi,\lambda) = \sum_{m_1,m_2}\rho_{m_1, m_2}(\varphi,\lambda) \cos^{m_1}[P_1(\varphi)] \sin^{m_1}[P_2(\varphi)]   \, , 
	\end{equation}
	with $m_1, m_2 \in \mathbb{Z}$ and $P_1(\varphi)$, $P_2(\varphi)$ polynomials in $\varphi^a$, \emph{provided} that $P_1(\varphi)$, $P_2(\varphi)$ are defined over compact domains only.
\end{itemize}

Now, assume to consider a supervised learning problem within a Quantum Gravity effective theory.
The model that the neural network ought to learn is either directly identified with a coupling that appears in the effective theory, or it is a quantity that is strictly related to some of these couplings -- see, for example, the learning problems mentioned in the introduction.
Thus, given the Tameness Conjecture, we should expect that the model that describes how the output depends on the input data is definable in a certain o-minimal structure.
Therefore, we shall make the following claim:
\begin{claimbox}
	In addressing neural network learning problems within an effective field theory of Quantum Gravity, the input space $X$, the output space $Y$ and the parameter space $\Omega$ are definable in a given o-minimal structure.
	Moreover, the family of functions
	\begin{equation}
		\label{o-min_function}
		f(x;\omega): \qquad X \times \Omega \rightarrow Y\,,
	\end{equation}
	among which the estimated model $f(x;\hat\omega)$ resides, must be definable in that o-minimal structure.
\end{claimbox}
In the next section, we will explore the implications of this claim for the learnability of a model formulated within a Quantum Gravity effective field theory.

%%%%%%%%%%%%%%%%%%%%%%%%%%%%%%%%%%%%%%%%%%%%%%%
%%%%%%%%%%%                 										%%%%%%%%%%%%%%%%%%%
%%%%%%%%  LEARNABILITY IN QUANTUM GRAVITY 		%%%%%%%%%%%%%%%%
%%%%%%%%%%%                 										%%%%%%%%%%%%%%%%%%%
%%%%%%%%%%%%%%%%%%%%%%%%%%%%%%%%%%%%%%%%%%%%%%%

\section{Learnability in Quantum Gravity}
\label{sec:QG_Learnability}

In this section we present the main idea of this work: by combining the definitions and concepts introduced Sections~\ref{sec:Deep_Learning_Definition}, \ref{sec:VC_dimension}, and~\ref{sec:o-minimal}, we show that Quantum Gravity is naturally embedded with learnability properties.
We proceed in steps: firstly, we describe how the finiteness of the shattering dimensions introduced in Section~\ref{sec:VC_dimension} reflects on the learnability of an algorithm; then, we illustrate which properties the shattering dimensions obey in o-minimal structures; finally, we show how Quantum Gravity naturally realizes neural network learnability, due to its tame, o-minimal nature.

\subsection{Relating learnability and finiteness of shattering dimensions}
\label{sec:QG_Learnability_VC}

Consider a family of functions $\mathcal{F}$: how can we tell whether such a family is \emph{learnable}, in the sense of the definitions delivered in Section~\ref{sec:Deep_Learning_Definition}? 
If the family of functions $\mathcal{F}$ is a set composed of a finite number of different functions, several theorems are available using which one can show that $\mathcal{F}$ is learnable, regardless of whether the functions within $\mathcal{F}$ ought to solve a classification or regression problem \cite{anthony_bartlett_1999}.

However, the case for which $\mathcal{F}$ is a set of finite cardinality is too restrictive. 
Indeed, typically, one would like $\mathcal{F}$ to include families comprising infinite functions. 
For instance, if we are facing a linear regression problem with a one-dimensional input space, $\mathcal{F}$ is composed of lines in $\mathbb{R}^2$, $f(x; w, \theta)$, depending on the slope $w$ and the intercept $\theta$.
Since both the slope $w$ and the intercept $\theta$ are generic real parameters, $\mathcal{F}$ is composed of an infinite set of lines.

In statistical learning theory, the requirement of finiteness of the cardinality of $\mathcal{F}$ to guarantee the learnability of $\mathcal{F}$ can be promoted to a requirement of finiteness of the shattering dimensions introduced in Section~\ref{sec:VC_dimension}, which can be adapted to the case where $\mathcal{F}$ is of infinite cardinality.
Below, we briefly recall some pivotal theorems that relate the learnability of $\mathcal{F}$ to the finiteness of the shattering dimensions for binary classifications and for general regressions.

\noindent\textcolor{colorloc3}{\textbf{Binary classification.}} The learnability of a set of functions $\mathcal{F}$ employed to address a binary classification problem is related to the finiteness of the Vapnik-Chervonenkis dimension of $\mathcal{F}$, as established by the following, important theorem \cite[Theorem 5.5]{anthony_bartlett_1999}:
\begin{importantbox}
	A family of functions $\mathcal{F}$ with domain in $X \subset \mathbb{R}^n$ and with values in $\{0,1\}$ is \emph{learnable} if and only if $\mathcal{F}$ has \emph{finite} Vapnik-Chervonenkis dimension.
	Moreover, the family $\mathcal{F}$ is learnable if and only if there exist two constants $c_1, c_2 > 0$ such that the inherent sample complexity \eqref{DL_mF} is bounded as
	\begin{equation}
		\label{LearnQG_bin_mH}
		\frac{c_1}{\epsilon^2} \log \frac{1}{\delta} < m_{\mathcal{F}}(\epsilon,\delta) < \frac{c_2}{\epsilon^2} \log \frac{1}{\delta}\,,
	\end{equation}
	for sufficiently small error $\epsilon$ and confidence $\delta$.
\end{importantbox}
The theorem, via \eqref{LearnQG_bin_mH}, also delivers an estimation of how large the sample should be in order for the algorithm to deliver an estimation with an error at most equal to $\epsilon$ (with respect to the optimal error \eqref{DL_opt}) and confidence parameter $\delta$. 
As expected, obtaining an estimation that is closer to the optimal error with high probability -- so that $\epsilon$ and $\delta$ are smaller -- requires larger samples.
It is also worth mentioning that, oftentimes, the above theorem is referred to as `\emph{fundamental theorem of statistical learning}'.

\noindent\textcolor{colorloc3}{\textbf{General regression.}} As stressed in Section~\ref{sec:VC_dimension}, the Vapnik-Chervonenkis dimension is not fit for analyzing the shattering of real-valued functions and, thus, the learning theorem given above cannot be directly applied to functions with real outputs. 
However, an analogous learning theorem can be formulated, which employs the fat-shattering dimension introduced in Section~\ref{sec:VC_fat-shattering} \cite[Theorem 19.1]{anthony_bartlett_1999}:
\begin{importantbox}
	Consider a family of functions $\mathcal{F}$ with domain $X \subset \mathbb{R}^n$ and with values in the unit interval $[0,1]$.
	We assume that the fat-shattering dimension of $\mathcal{F}$ is \emph{finite}, namely $\text{fat}_\mathcal{F}(\gamma) < \infty$ for any $\gamma > 0$.
	
	Consider an approximate sample error minimization algorithm $\mathcal{A}$ for the family $\mathcal{F}$ and define $\ell(z) := \mathcal{A} \left(z, \frac{\epsilon_0}{6}\right)$ with $\epsilon_0 = \frac{16}{\sqrt{N_{\text{data}}}}$.
	Then, $\ell$ is a learning algorithm for $\mathcal{F}$ with sample complexity \eqref{DL_mL} bounded as
	\begin{equation}
		\label{LearnQG_gen_mLb}
		m_{\ell} (\epsilon, \delta) \leq m_0(\epsilon, \delta) = \frac{256}{\epsilon^2} \left[ 18\, \text{fat}_{\mathcal{F}}\left( \frac{\epsilon}{256} \right) \log^2\left( \frac{128}{\epsilon} \right) + \log\left( \frac{16}{\delta} \right) \right]
	\end{equation}
	for any error $\epsilon > 0$ and confidence $\delta > 0$.
\end{importantbox}

As noticed for the binary classification learning theorem stated above, the upper bound on the complexity grows when the error $\epsilon$ with respect to the optimal value or the confidence parameter $\delta$ decreases.
Moreover, the algorithm $\ell$ that learns the model is defined out of an approximate sample error minimization algorithm that obeys \eqref{DL_approx_sem}, for a strict minimization of the error might be unfeasible.

\subsection{Shattering in o-minimal structures}
\label{sec:QG_Learnability_o-min}

Above, we have been agnostic about the functions that ought to compose the family $\mathcal{F}$ among which the model sought by the algorithm resides, for we have only focused on the kind of output they should deliver.
However, in Section~\ref{sec:o-minimal_Quantum_Gravity} we have shown that Quantum Gravity effective theories are characterized by couplings and interactions that are definable within o-minimal structures, such as those listed at the end of Section~\ref{sec:o-minimal_Review}.
Then, as postulated at the end of Section~\ref{sec:o-minimal_Quantum_Gravity}, we may assume that the family $\mathcal{F}$ to learn via a neural network algorithm must be definable in a given o-minimal structure.
The core of this section is to illustrate that this requirement has crucial consequences for the shattering dimensions of the family $\mathcal{F}$.
Again, we split the discussion into two cases, according to whether the functions populating the family $\mathcal{F}$ are binary- or real-valued.

\noindent\textcolor{colorloc3}{\textbf{Binary classification.}} Preliminarily, it is convenient to introduce \emph{definable families} of sets, that are defined as follows.
Consider an o-minimal structure defined over a set $M$. For any definable subset $S \subseteq M^{n+k}$ and $w \in M^k$, we identify the `\emph{fiber}' of the set $S$ at the point $w$ as
\begin{equation}
	\label{LearnQG_o-min_fam}
	S_w = \{ v \in M^n \; :\;  (v,w) \in S \}\,.
\end{equation}
Then, a \emph{definable family} is the collection $(S_w)_{w\in M^k}$, for varying $w$.
In particular, given an o-minimal structure $\mathbb{R}_{\mathcal{A}}$, the set $S$ and any of its fibers $S_w$ stem from a locus constructed out of the functions in $\mathcal{A}$, and logical operators.

\begin{figure}[thb]
	\centering
	\includegraphics[height=7cm]{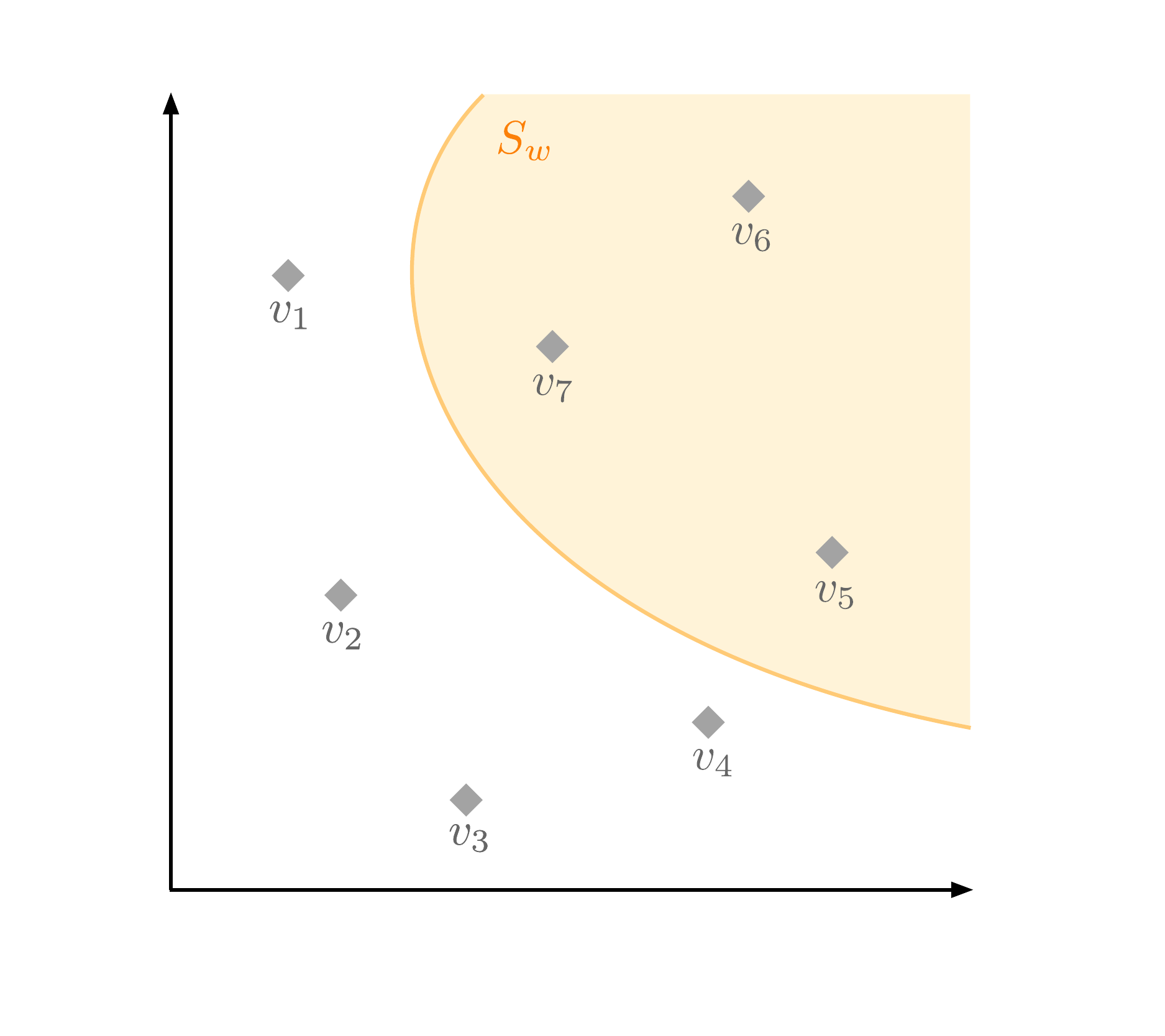}
	\caption{\footnotesize A two-dimensional example of a set of seven points $v_1, \ldots, v_7$ that is partitioned into two sets $\{v_1, v_2, v_3, v_4\}$ and $\{v_5, v_6, v_7\}$ via a fiber $S_w$ that contains only the latter subset.  One can then identify $S_w = S_1$, and $S_0$ as its complement. \label{Fig:LearnQG_VC}}
\end{figure}

Now, consider an arbitrary set of points $v_1, \ldots, v_m \in M^n$, and assume that each of them is labeled either with $0$ or $1$ -- this may be considered a `data set' for a binary classification problem as in Section~\ref{sec:Deep_Learning_Definition_Framework}.
The fibers \eqref{LearnQG_o-min_fam} may then be employed to define a locus within $M^n$ separating the variables $v$ into two subsets, say $S_0$ and $S_1$ -- see Figure~\ref{Fig:LearnQG_VC}.
Such a separation may then allow us to label the points within each set such that, if $v \in S_0$, then its label is $0$ and, if $v \in S_1$, its label is $1$.
How such a partition into two subsets is performed depends on the parameters $w$.
Ideally, we would like that all the points among $v_1, \ldots, v_m$ that are labeled with $0$ lie inside the set $S_0$ and those labeled with $1$ fall into the set $S_1$.
In sum, $S_w$ ought to be able to shatter all the possible $2^m$ subsets of the set of points $v_1, \ldots, v_m$, if we wish to be sure that any of the labeling assignments are realized with the definable family $S_w$.

In o-minimal structures, the above partitioning \emph{cannot} always be realized, as this is captured by the following, critical property that every o-minimal structure enjoys:
\begin{importantbox}
	In every o-minimal structure, any definable family $(S_w)_{w\in M^k}$ has finite Vapnik-Chervonenkis dimension.
\end{importantbox}
This result, proved in \cite{Laskowski1992VapnikChervonenkisCO}, delivers a different viewpoint on how o-minimal structures are `\emph{tame}'.
Furthermore, it should be expected from the general discussion of o-minimal structures of Section~\ref{sec:o-minimal}: in order, for a fiber $S_w$, to correctly label any infinite set of points $v_1, v_2, \ldots \in M^n$ is to use a fiber $S_w$ defined via a function that has infinite `turning points'.
An example would be a fiber defined via a periodic function, but these functions cannot belong to o-minimal structures, for their projections may contain sets with infinite disconnected components.

It is worth stressing that one can show that the crucial property stated above is grounded in the model theory definition of o-minimal structures.
In fact, every o-minimal structure is equipped with the \emph{non-independence property} from which the above finiteness property follows \cite{Tressl:2010}. 
In Appendix~\ref{sec:MT_o-min_NIP} we review how o-minimal structures are defined in model theory, and state their \emph{non-independence property}.

\noindent\textcolor{colorloc3}{\textbf{General regression.}} General regression problems -- and multi-class classification problems as subcases thereof -- are more subtle to investigate.
Indeed, as highlighted in Section~\ref{sec:VC_fat-shattering}, the Vapnik-Chervonenkis dimension is not fit for analyzing the learnability of real-valued functions, for one should rather consider the fat-shattering dimension therein defined.

The determination of whether the fat-shattering dimension is finite for o-minimal structures relies on the same definable families, defined out of the fibers \eqref{LearnQG_o-min_fam}, that we already employed for the binary classification problems.
In particular, if the fibers \eqref{LearnQG_o-min_fam} can fat-shatter any arbitrarily large set of points $v_1, \ldots, v_n \in M^n$ and for any real number $\gamma>0$ determining the fat-shattering, then the definable family $(S_w)_{w\in M^k}$ would be characterized by infinite fat-shattering dimension.

\begin{figure}[thb]
	\centering
	\includegraphics[height=6cm]{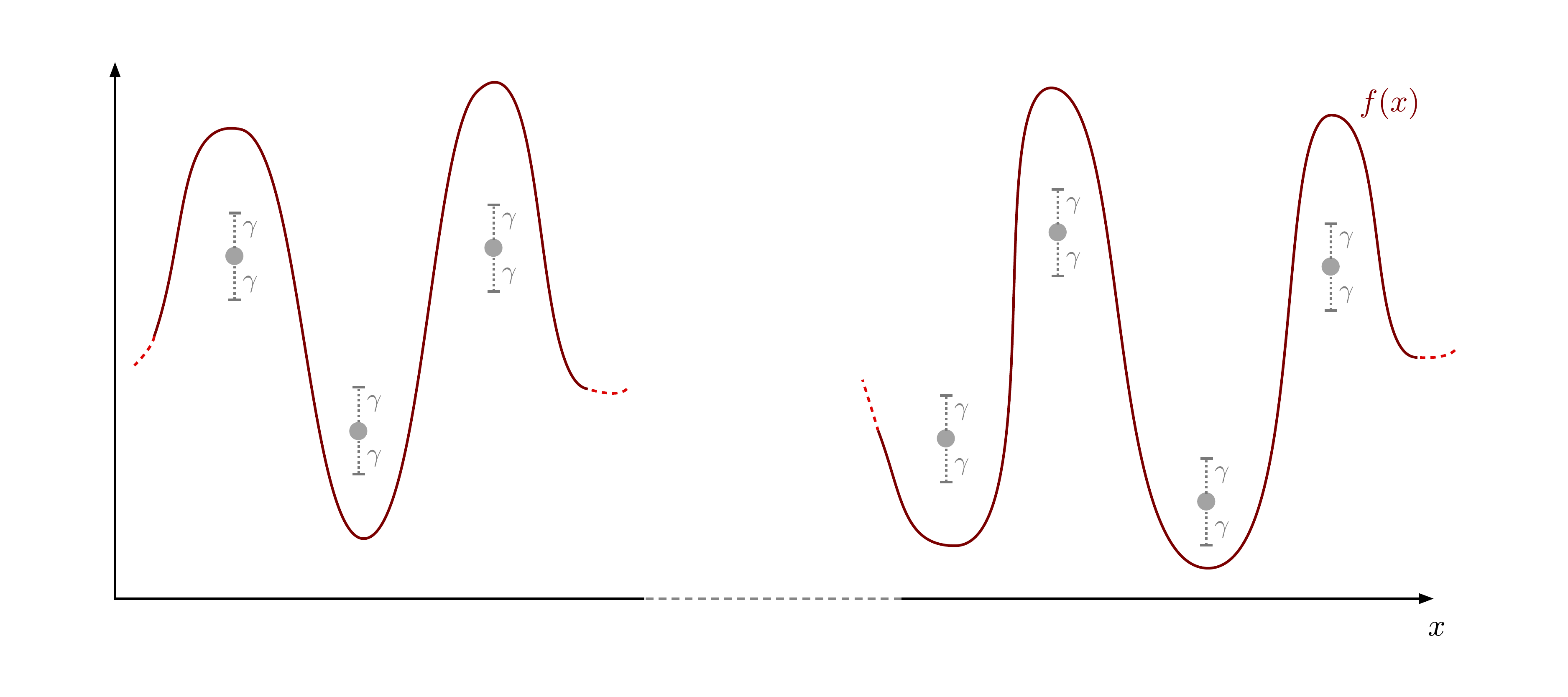}
	\caption{\footnotesize An example of a function $f(x)$, with the real line as domain, with infinite fat-shattering dimension. Since the function needs to have an infinite number of turning points, it does not have regular tails and thus it cannot satisfy the Tame monotonicity theorem stated in Section~\ref{sec:o-minimal_Review}.  \label{Fig:LearnQG_fat-shatt}}
\end{figure}

Unfortunately, unlike the Vapnik-Chervonenkis dimension, there is presently no mathematical theorem that guarantees that every definable family defined in any o-minimal structure is characterized by finite fat-shattering dimension.
However, it is not expected that the fat-shattering dimension of a definable family is infinite.
In fact, recalling the very definition of fat-shattering given in Section~\ref{sec:VC_fat-shattering}, one sees that an infinite fat-shattering dimension would imply that the fibers $S_w$ can separate sets composed of \emph{infinite} disconnected components, in any possible subset and at any arbitrary separation as dictated by $\gamma$.
Hence, given an o-minimal structure $\mathbb{R}_{\mathcal{A}}$, the functions in $\mathcal{A}$ that define these fibers should have infinite `turning points' (or `turning loci') -- see Figure~\ref{Fig:LearnQG_fat-shatt} for a pictorial representation of the problem. 
In turn, this would imply that there exist some projections of these fibers constituted by an infinite number of disconnected components -- for instance, the extrema of the functions defining the fibers -- thus invalidating the o-minimality assumption.

Therefore, analogously to the finiteness of the Vapnik-Chervonenkis dimension, we shall assume the following:
\begin{importantbox}
	In the o-minimal structures under consideration, any definable family $(S_w)_{w\in M^k}$ is characterized by a finite fat-shattering dimension.
\end{importantbox}
In the next section we will illustrate the implications of this assumption for the learnability in Quantum Gravity effective theories.

\subsection{Learnability in Quantum Gravity effective field theories}
\label{sec:QG_Learnability_EFTs}

Finally, we are in the position to show the learnability within effective field theories that admit an ultraviolet completion within Quantum Gravity.
As stressed in Section~\ref{sec:o-minimal_Quantum_Gravity}, the key observation is that the interactions and couplings of any low-energy effective field theory have to be definable in a certain o-minimal structure, and these structures are characterized by finite shattering dimensions.
Thus, due to the results of Section~\ref{sec:QG_Learnability_VC}, learning algorithms exist that are capable of learning functions defined in these structures.

Indeed, recall, from Section~\ref{sec:o-minimal_Quantum_Gravity}, that the definable sets, using which the tame structures of Quantum Gravity effective theories are defined, have a fibered structure: for each set of parameters $\lambda^\kappa$, generically belonging to a subset of $\mathcal{P} \subseteq \mathbb{R}^k$, the moduli $\varphi^a$ belong to a moduli space patch $\mathcal{M}^{\text{loc}}_\lambda \subseteq \mathbb{R}^N$.
Therefore, the definable set may be regarded as the unions of a definable family of the sets
\begin{equation}
	\label{LearnQG_QG_Sl}
	S_\lambda = \{ \varphi \in \mathcal{M}^{\text{loc}}_\lambda \; :\;  (\varphi, \lambda) \in \mathcal{M}^{\text{loc}}_\lambda \times \mathcal{P} \}\,.
\end{equation}
The fibers just defined share the same structure as those defined in \eqref{LearnQG_o-min_fam}, and they inherit their properties: the definable family of fibers \eqref{LearnQG_QG_Sl}, $(S_\lambda)_{\lambda \in \mathcal{P}}$ is characterized by finite Vapnik-Chervonenkis dimension and is expected to have finite fat-shattering dimension as well.
Moreover, due to the projection properties that o-minimal structures enjoy, any subset of these fibers is also definable, and characterized by the same finite shattering dimensions.

Indeed, the functions that a neural network-based algorithm ought to learn may be defined over those fibers, or subsets thereof. 
For instance, let us consider a regression problem, such as the one mentioned in the introduction, of finding out the behavior of a real coupling $g(\varphi;\lambda)$ across the moduli space $\mathcal{M}_\lambda$. 
Generically, the coupling is a function
\begin{equation}
	\label{LearnQG_QG_g}
	g(\varphi ; \lambda): \qquad \mathcal{M}_\lambda \times \mathcal{P} \rightarrow \mathbb{R}\,.
\end{equation}
We assume that we do know which values $g_i$ the coupling must take at some moduli space points $\varphi_i^a$, and we wish to estimate which parameters $\hat{\lambda}^\kappa$ best model these known values; namely, we wish to find the set of parameters $\hat{\lambda}^\kappa$ so that $g(\varphi_i ; \hat{\lambda})$ are sufficiently close to the given values $g_i$.
Thus, the input space may be characterized by a local patch the moduli space $\mathcal{M}_\lambda$, the output space is (a subset of) the real line, and the parameter space to be searched by the algorithm coincides with the geometrical parameter space spanned by $\lambda^\kappa$.
Therefore, as per the results of Section~\ref{sec:QG_Learnability_VC}, the finiteness of the fat-shattering dimension of the fibers \eqref{LearnQG_QG_Sl} implies the learnability of the function $g(\varphi ; \hat{\lambda})$, provided that the sample of training data points is large enough.
Clearly, a similar conclusion holds if some of the moduli or some of the internal geometrical parameters are fixed, or if one has to find a function of the parameters, rather than of the moduli.

Alternatively, consider the binary classification problem mentioned in the introduction about determining whether an effective theory may accommodate slow-roll inflation.
We may assume that the moduli space point is fixed, so that $\mathcal{M}_\lambda$ collapses to a single point, the same for every value of the parameters $\lambda$.
Then, at this single moduli space point, we may determine whether the slow-roll conditions for realizing inflation are satisfied. 
Here, the input space is given by a finite set of parameters $\lambda^\kappa_i$, at which we do know whether the slow-roll conditions are met, whereas the output space is the binary set $\{0,1\}$, encoding the fact that, for the parameters $\lambda^\kappa$, slow-roll conditions are locally satisfied or not satisfied.
We can thus introduce the boolean family of functions
\begin{equation}
	\label{LearnQG_QG_infl}
	B(\lambda; \omega): \qquad \mathcal{P} \times \Omega \rightarrow \{0,1\}\,,
\end{equation}
with the values $0$ and $1$.
The parameters $\omega$ that specify the shape of the function are, in this case, external to the tame set $\mathcal{M}_\lambda \times \mathcal{P}$.
However, since the function in \eqref{LearnQG_QG_infl} is essentially derived out of the slow-roll conditions, which are in turn defined out of quantities appearing explicitly in the Quantum Gravity low-energy theory, one should expect the function \eqref{LearnQG_QG_infl} to be definable as well; subsequently, also the parameter space $\Omega$ has to be definable.
Hence, the functions of the family \eqref{LearnQG_QG_infl} lead to sets of finite Vapnik-Chervonenkis dimension, and one can find a learning algorithm to get the function $B(\lambda; \hat\omega)$ that best describes the training set.

These observations can be extended to larger classes of learning problems within Quantum Gravity effective theories.
Indeed, employing the learnability results of Section~\ref{sec:QG_Learnability_VC}, and given the shattering dimension properties of any o-minimal structures outlined in Section~\ref{sec:QG_Learnability_o-min}, we can then make the following, general statement:
\begin{claimbox}
	In every low-energy effective theory of Quantum Gravity, the definability of the interactions in a given o-minimal structure implies that there exist learning algorithms that learn the interactions, or any other tame quantity defined out of those. 
	In particular, any classification problem or regression problem that involves the tame interactions that appear in the low-energy theory is learnable.
\end{claimbox}
It is worth stressing that our findings concern learnability, and not \emph{decidability}. 
Indeed, unlike learnability, proving the decidability in a given o-minimal structure is a hard task, as outlined in Appendix~\ref{sec:MT_o-min_dec}.
For instance, there is no proof of decidability within the simple $\mathbb{R}_{\text{exp}}$ o-minimal structure, or for the more complicated o-minimal structures, such as $\mathbb{R}_{\text{an,exp}}$, which are relevant for Quantum Gravity effective field theories.

%%%%%%%%%%%%%%%%%%%%%%%%%%%%%%%%%%%%%%%%%%%%%%%
%%%%%%%%%%%                 										%%%%%%%%%%%%%%%%%%%
%%%%%%%%  							CONCLUSIONS						%%%%%%%%%%%%%%%%
%%%%%%%%%%%                 										%%%%%%%%%%%%%%%%%%%
%%%%%%%%%%%%%%%%%%%%%%%%%%%%%%%%%%%%%%%%%%%%%%%

\section{Conclusions and outlook}
\label{sec:Conclusions}

In this work we have shown that effective field theories originating from compactifications of string theory are naturally endowed with learnability properties, in the statistical meaning described in Section~\ref{sec:Deep_Learning_Definition}.
The learnability of such effective field theories is grounded in their underlying geometry: any coupling, interaction or physical quantity in consistent Quantum Gravity effective field theories must be definable within a certain o-minimal structure.
Our findings imply that large classes of problems that can be addressed within string theory effective field theories can be learned, including binary or multi-class classification problems and regression or interpolation problems.

The results of this work are general, and eventually can be checked in some concrete scenarios.
In particular, the learnability theorems listed in Section~\ref{sec:QG_Learnability_VC} carry information about the sample complexity that is required for learning an algorithm within given error and confidence.
It would be interesting to compute the sample complexity in concrete applications, with common gradient-based optimization algorithms.

It would be also interesting to employ our findings to substantiate the learnability of some properties of the string landscape.
For example, one could employ our results to infer whether a given conjectural property that theories in the landscape should possess -- such as those suggested within the Swampland program -- can be learned via an algorithm.
Or, more generally, it would be tantalizing to employ our statements in order to establish a catalog of problems within the string landscape that can be learned.
A direct investigation in this direction is left for future work.

\vspace{1cm}

\noindent\textcolor{colorloc3}{\textbf{Acknowledgments.}} I am deeply grateful to Thomas Grimm, Luca Martucci, Mick van Vliet, Timo Weigand, and Alexander Westphal for valuable discussions and comments on the manuscript.
This research is supported in part by Deutsche Forschungsgemeinschaft under Germany’s Excellence Strategy EXC 2121 Quantum Universe 390833306 and by Deutsche Forschungsgemeinschaft through a German-Israeli Project Cooperation (DIP) grant “Holography and the Swampland”.

\appendix

\section{o-minimality, independence property and decidability}
\label{sec:MT_o-min}

In this section we overview some basic facts of model theory that support the conclusions threaded in Section~\ref{sec:QG_Learnability_o-min}.
Here, for the ease of the exposition, we will be qualitative, and we refer to \cite{Marker2002,Moerdijk2018} for a precise, mathematical treatment of the subjects presented in this section.

\subsection{Independence property in o-minimal structures}
\label{sec:MT_o-min_NIP}

As mentioned in Section~\ref{sec:o-minimal}, o-minimal structures are objects that were firstly introduced in model theory.
A key property that they are endowed with is the \emph{not independence property}, from which the findings of Section~\ref{sec:QG_Learnability_o-min} follow.
In order to explain this property, we will recall some basic definitions and facts of model theory, in the framework of first-order logic.

In logic, mathematical properties can be framed within \emph{languages}.
A language is is composed of three sets of symbols: constants, functions, and relations; in particular, functions and relations may take an arbitrary number of symbols as arguments. 
Within a language, one may introduce \emph{auxiliary symbols}: these can include variables $v_1, v_2, v_3, \ldots$, logical operators such as the equality symbol $=$, the absurdity symbol $\perp$, the Boolean connectives -- the conjunction `and' $\wedge$, the disjunction `or' $\vee$, the negation $\lnot$, the implication `if ... then' $\Rightarrow$ -- and the quantifiers -- the existential one $\exists$ and the universal one $\forall$.

Out of the symbols defining the language and the auxiliary symbols, one can introduce \emph{formulas} to state properties.
Within a formula, a variable may appear within a quantifier, $\exists$ or $\forall$: if it is the case, we say that the variable is \emph{bound}, otherwise we say that it is \emph{free}.
A formula -- that we denote as $\phi$ -- that has no free variable is called a \emph{sentence}.
Sentences play a crucial role in logic, for they can be used to state general properties that variables ought to obey.
Indeed, a set of sentences constitute a \emph{theory} $\mathscr{T}$, and the \emph{axioms} of a theory are its defining sentences.

Languages, and theories defined within them are abstract objects. 
However, languages may be concretely realized within a given set, called \emph{universe} or \emph{domain}, and these realizations go by the name of \emph{structures}.
A theory formulated within a language that is realized within the universe set is called \emph{model}.
Namely, we say that $\mathscr{M}$ is a model for the theory $\mathscr{T}$ is every sentence $\phi \in \mathscr{T}$ is realized in $\mathscr{M}$. 
Moreover, if a sentence is realized within a model $\mathscr{M}$, we write $\mathscr{M} \models \phi$.

As an example, consider the language of ordered rings $\mathscr{L}_{\text{or}}$. This is constituted by the binary functions $+$, $-$, $\cdot$ and the binary relation $<$ and the constant symbols $1$ and $0$. 
Example of formulas include $v_1 = 0 \vee v_1 > 0$, or $\exists v_2 \; v_2 \cdot v_2 = v_1$, for some variables $v_1$ and $v_2$; the first formula states that $v_1 \geq 0$, while the second that $v_1$ is a square, and in both cases there is a free variable.
An example of a sentence, where there are no free variables, is $\forall v_1\; ( v_1 \neq 0 \Rightarrow \exists v_1 \; v_1 \cdot v_2 = 1)$, which states that a variable has an inverse (with respect to the function $\cdot$), whenever it is different from zero.
A structure that realizes this language is given, for instance, by the set of integers $\mathbb{Z}$, equipped with the usual symbols; or it can be realized within the set of rational numbers $\mathbb{Q}$. 
Notice, however, that in the latter structure the above sentence is satisfied, unlike in the former.
 
Consider a model $\mathscr{M}$ for a given structure defined over a set $M^{n+m}$.
A set $S \subseteq M^n$ is said to be \emph{definable} if and only if there exists a formula $\phi (v,w)$, with $v \in \mathbb{M}^n$ and $w \in \mathbb{M}^m$ such that the set $S$ can be written as
\begin{equation}
	\label{MT_o-min_def}
	S =\{ v \in M^n \; :\; \phi (v,w) = 0 \}\,, 
\end{equation}
and the formula $\phi (v,w)$ is said to \emph{define} the set $S$.

For instance, consider the model of ordered rings defined over the real set $\mathbb{R}$. Examples of definable sets on the real line include the set of zeroes of polynomials:
\begin{equation}
	\label{MT_o-min_pol}
	S =\{ x \in \mathbb{R} \; :\; P(x) = 0 \}\,, \quad \text{with $P(x) = \sum\limits_{i = 0}^{m-1} a_i x^i$}\,,
\end{equation}
and the formula defining the set is simply given by $\phi(x, a_0, \ldots, a_{m-1}) := P(x) = 0$.

A structure on which an order is defined may constitute an \emph{o(rder)-minimal structure}, whenever the following holds:
\begin{importantbox}
	An ordered structure $(M, <, \ldots)$ is called \emph{o-minimal} if, given a definable set $S \subset M$, there exist finitely many intervals $I_1, \ldots, I_k$ with endpoints in $M \cup \{ \pm \infty \}$ and a finite set $S_0$ such that the set $S$ can be written as
	\begin{equation}
		\label{MT_o-min_def_b}
		S = S_0 \cup I_1 \cup \ldots I_k \,.
	\end{equation}
\end{importantbox} 
Indeed, the definition of o-minimal structures given in Section~\ref{sec:o-minimal_Review} can be traced to the above, fundamental model theoretical definition of o-minimal structures, and the loci \eqref{o-min_Loci} are examples of definable sets within these structures, with the polynomials $P_i$ and $Q_j$ constituting the formulas that define these sets.

Furthermore, as done in Section~\ref{sec:QG_Learnability_o-min}, it is convenient to introduce a \emph{definable family} of sets.
Consider a definable set $S \subseteq M^{n+k}$ and $w \in M^k$; we define
\begin{equation}
	\label{MT_o-min_fam}
	S_w = \{ v \in M^n \; :\;  (v,w) \in S \}\,, 
\end{equation}
which can be thought as being a `\emph{fiber}' of the set $S$ above $w$. 
Then, a \emph{definable family} is $(S_w)_{w \in M^k}$. Eventually, the set $S$ can be considered a definable family of sets.

Equipped with the definition of definable family, we can finally state the \emph{independence property}.
Consider a formula $\phi(v,w)$ defined within the theory $\mathscr{T}$. The formula $\phi(v,w)$ has the \emph{independence property} if, in every model $\mathscr{M}$ of the theory $\mathscr{T}$, the definable family $(S_w)_{w \in M^k}$ defined by $\phi(v,w)$ shatters an infinite subset of $M^n$.
Namely, there are $v_i \in M^n$ and $w_I \in M^k$ such that
\begin{equation}
	\label{MT_o-min_ind}
	\mathscr{M} \models \phi(v_i , w_I) \quad \Leftrightarrow \quad i \in I\,,
\end{equation}
where, $i = 1, \ldots, n$ denotes the index of the tuple $v_i$ and $I$ is any subset of these indices.
Said differently, consider an arbitrary set of variables, $v_1, \ldots, v_m$, with $m$ arbitrary; then, \eqref{MT_o-min_ind} tells that, for any subset of these variables, there is a formula, specified by the parameters $w_I$, that renders the subset of these variables definable.
In turn, this means that the formula $\phi(v, w)$ defines \emph{all} the subsets of any arbitrary number of variables; namely a formula is equipped with the independence property if and only if the associated definable family of sets is of infinite Vapnik-Chervonenkis dimension.

If \emph{none} of the formulas within a theory are equipped with the independence property, then we say that the theory is endowed with the \emph{non-independence property}.
One can show that \emph{every} o-minimal structure is characterized by the non-independence property \cite{Laskowski1992VapnikChervonenkisCO,Tressl:2010}.
As such, any o-minimally definable family is necessarily constituted by finite Vapnik-Chervonenkis dimension, a result upon which several of the conclusions reported in Sectio~\ref{sec:QG_Learnability_o-min} rely.

\subsection{Learnability and decidability}
\label{sec:MT_o-min_dec}

In Section~\ref{sec:Deep_Learning_Definition} we have defined what we mean with `\emph{learnability}', using which the results of Section~\ref{sec:QG_Learnability} are formulated.
However, model theory is endowed with a different notion, the one of \emph{decidability}:
a theory is \emph{decidable} if, for every sentence $\phi$, there is an algorithm such that, given $\phi$ as input, it is able to tell whether $\phi$ is satisfied within the theory.

Although the two definitions might look similar, the difference between them is crucial.
In fact, the learnability definitions delivered in Section~\ref{sec:Deep_Learning_Definition} have a statistical nature.
Consider, for instance, a regression model and assume that, with an algorithm, we are able to get a function that model our data. 
Firstly, the original data set may not be modeled with exact precision, for some predicted outputs may well be different from the actual output of the training data set.
Secondly, any predicted output is associated with a probability estimation; namely, an output is predicted to have a certain value with a given probability, related to the confidence parameters.

On the contrary, decidability is more stringent and general, and generically it does not have a statistical nature.
Furthermore, while it is simple to show that o-minimal structures are endowed with learnability, as we did in Section~\ref{sec:QG_Learnability}, proving the decibility of a given o-minimal structure is a hard task. 
For instance, the decidability of the o-minimal structure $\mathbb{R}_{\text{alg}}$ is a consequence of the Tarski–Seidenberg theorem \cite{Tarski:1951,Seidenberg:1954} (see also \cite{van-den-Dries:1988}).
However, the decidability of the $\mathbb{R}_{\text{exp}}$ structure has not been established yet, for its proof relies on the usage of Schanuel's conjecture \cite{Macintyre1996-MACOTD-3}.

%\bibliographystyle{jhep}
%\bibliography{references.bib}

\providecommand{\href}[2]{#2}\begingroup\raggedright\endgroup

\end{document}